\documentclass[aps,prc,twocolumn,showpacs,superscriptaddress]{revtex4}

\usepackage{amsmath}
\usepackage{graphicx,color}

\newcommand{\pt}{p_T}
\newcommand{\nn}{\nonumber\\}
\newcommand{\ba}{\begin{eqnarray}}
\newcommand{\ea}{\end{eqnarray}}
\newcommand{\la}[1]{\label{#1}}

\def\lsim{\,\raise0.3ex\hbox{$<$\kern-0.75em\raise-1.1ex\hbox{$\sim$}}\,}
\def\gsim{\,\raise0.3ex\hbox{$>$\kern-0.75em\raise-1.1ex\hbox{$\sim$}}\,}

\begin{document}

\title{Multiple jets and $\gamma$-jet correlation in high-energy heavy-ion collisions}


\author{Tan Luo}
\affiliation{Key Laboratory of Quark and Lepton Physics (MOE) and Institute of Particle Physics, Central China Normal University, Wuhan 430079, China}
\author{Shanshan Cao}
\affiliation{Department of Physics and Astronomy, Wayne State University, Detroit, Michigan 48201}
\author{Yayun He}
\affiliation{Key Laboratory of Quark and Lepton Physics (MOE) and Institute of Particle Physics, Central China Normal University, Wuhan 430079, China}
\author{Xin-Nian Wang}
\affiliation{Key Laboratory of Quark and Lepton Physics (MOE) and Institute of Particle Physics, Central China Normal University, Wuhan 430079, China}
\affiliation{Nuclear Science Division Mailstop 70R0319,  Lawrence Berkeley National Laboratory, Berkeley, CA 94740}

\begin{abstract}
$\gamma$-jet production is considered one of the best probes of the hot quark-gluon plasma in high-energy heavy-ion collisions since the direct $\gamma$ can be used to gauge the initial  energy and momentum of the associated jet. This is investigated within the Linear Boltzmann Transport (LBT) model for jet propagation and jet-induced medium excitation. With both parton energy loss and medium response from jet-medium interaction included, LBT can describe experimental data well on $\gamma$-jet correlation in Pb+Pb collisions at the Large Hadron Collider. Multiple jets associated with direct $\gamma$ production are found to contribute significantly to $\gamma$-jet correlation at small $p_T^{\rm jet}<p_T^\gamma$ and large azimuthal angle relative to the opposite direction of $\gamma$. Jet medium interaction not only suppresses the leading jet at large $p_T^{\rm jet}$ but also sub-leading jets at large azimuthal angle. This effectively leads to the narrowing of $\gamma$-jet correlation in azimuthal angle instead of broadening due to jet-medium interaction.  The $\gamma$-jet profile on the other hand
will be broadened due to jet-medium interaction and jet-induced medium response. Energy flow measurements relative to the direct photon is illustrated to reflect well
the broadening and jet-induced medium response.
\end{abstract}


\maketitle

\section{Introduction}
Parton energy loss during jet propagation in dense medium can lead to suppression of the single inclusive hadron spectra at large transverse 
momentum \cite{Wang:1991xy,Wang:1998ww,Qin:2007rn,Chen:2011vt,Buzzatti:2011vt,Majumder:2011uk,Adcox:2001jp,Adler:2002xw,Aamodt:2010jd,CMS:2012aa}, 
large transverse momentum $\gamma$-hadron \cite{Wang:1996yh,Renk:2006qg,Zhang:2009rn,Qin:2009bk,Adare:2009vd,Abelev:2009gu} and dihadron 
correlations \cite{Zhang:2007ja,stardihadron} in high-energy heavy-ion collisions.  Phenomenological study of experimental data on these observables at the Relativistic Heavy-ion Collider (RHIC) and the Large Hadron Collider (LHC) have provided important constraints on the jet transport coefficient in high-energy heavy-ion collisions \cite{Burke:2013yra}. Production and suppression of fully reconstructed single jets, di-jets and $\gamma$-jets have also been studied in heavy-ion 
collisions \cite{Vitev:2009rd,Qin:2010mn,Young:2011qx,He:2011pd,Renk:2012cx,Dai:2012am,Chien:2015hda,Kang:2017xnc,Aad:2010bu,Chatrchyan:2011sx,Chatrchyan:2012gt,Chatrchyan:2012nia,Abelev:2013kqa,Aad:2014bxa} and they can provide additional constraints on the jet-medium interaction and jet transport coefficient.

The jet transport coefficient $\hat q$ is related to the gluon distribution density \cite{CasalderreySolana:2007sw} of the medium at the scale of averaged transverse momentum transfer between the propagating jet and thermal medium partons. It can be defined alternatively as the average transverse momentum broadening squared of a propagating parton per unit length. One can therefore in principle measure jet transport coefficient directly through di-hadrdon, $\gamma$-hadron, di-jet or $\gamma$-jet correlation in azimuthal 
angle \cite{Appel:1985dq,Blaizot:1986ma}. Though the Sudakov form factor from initial state radiation dominates the azimuthal angle correlation of di-jets and $\gamma$-jets with large transverse momentum, especially at LHC \cite{Mueller:2016gko,Chen:2016vem}, it has been proposed that the large angle correlation could be influenced by large angle scattering between jet shower and medium partons that can provide insight into the emergence of strongly interacting fluid from an asymptotically weakly interaction theory \cite{DEramo:2012uzl}.

In this paper, we will study $\gamma$-jet correlation within the framework of Linear Boltzmann Transport (LBT) Monte Carlo 
model \cite{Li:2010ts,Wang:2013cia,He:2015pra,Cao:2016gvr,Cao:2017hhk} for jet production and propagation in heavy-ion collisions. We will in particular focus on the effect of multiple jet production and suppression in the $\gamma$-jet correlation. Multiple jets are produced from the large angle radiative processes in the initial
hard processes and their effects have been studied  in jet quenching and multiple hadron correlation \cite{Ayala:2009fe,Ayala:2011ii,Ayala:2015jaa}.  The fractional contributions from multiple jets to $\gamma$-jet correlation are on the average in the order of the strong coupling constant $\alpha_{\rm s}$. However, their contributions can become significant and even dominant in the region of large momentum imbalance $p_T^{\rm jet}<p_T^\gamma$ and large azimuthal angle difference $|\phi^\gamma-\phi^{\rm jet}-\pi|$. Energy loss and suppression of these sub-leading jets can lead to medium modification of the $\gamma$-jet correlation in these kinematic regions in addition to the modification caused by energy loss and suppression of the leading jet in $\gamma$-jet events.

When jet partons propagate through the quark-gluon plasma, jet-medium interactions will lead to the reduction of the final jet energy. In addition, jet-medium interaction will also lead to the medium excitation and redistribution of the lost energy lost inside and outside the jet cone because of the further propagation of recoil partons. These recoil partons from jet-medium interaction are very important and should be taken into account in the final jet 
reconstruction~\cite{Li:2010ts,Wang:2013cia,He:2015pra,Casalderrey-Solana:2014bpa,Tachibana:2015qxa,Wang:2016fds,Tachibana:2017syd,KunnawalkamElayavalli:2017hxo,Milhano:2017nzm}. Significant modifications of the jet structure have been found in recent experimental data on single inclusive jets in Pb+Pb collisions at the LHC \cite{Chatrchyan:2013kwa,Chatrchyan:2014ava,Aaboud:2017bzv} that provides strong evidence for jet-induced medium excitation and the redistribution of energy and momentum inside and outside the jet cone.  Recent data on the $\gamma$-jet fragmentation function \cite{Sirunyan:2018qec} also show strong enhancement of soft hadrons from jet-induced medium excitation inside the jet-cone as well as the suppression of leading hadrons due to parton energy loss.  In this paper, we will use LBT model to study $\gamma$-jet correlation and the effects of multiple jets. We will also carry out the baseline study of $\gamma$-jet correlation as a function of $p_T^{\rm jet}$ and $\gamma$-jet asymmetry variable $x_{J\gamma}$ and their modification due to jet propagation in medium and  contributions from medium recoil. We will investigate contributions from multiple jets to the $\gamma$-jet correlation and provide predictions from LBT on the transverse profile of $\gamma$-jet and energy flow due to jet propagation and medium response in high-energy heavy-ion collisions at LHC.

The remainder of the paper is organized as follows. After providing a brief introduction to the LBT model in Sec. \ref{LBT}, we will show in Sec. \ref{asym} results from LBT on $\gamma$-jet correlation in energy asymmetry, the transverse momentum of the associated jet and contributions from multiple jets. We will also show the averaged jet energy loss responsible for the modification of the $\gamma$-jet correlation and the effect of medium recoil or jet-induced medium excitation. The $\gamma$-jet correlation in azimuthal angle is investigated in Sec.~\ref{azym}. Multiple jets are shown to dominate in the large angle and energy loss of the sub-leading jets leads to the suppression of $\gamma$-jet correlation at large angles. In Sec.~\ref{flow}, we will examine the modification of $\gamma$-jet transverse profile and energy flow due to jet-medium interaction and jet-induced medium excitation. A summary and some discussions are given in Sec.~\ref{summary}.

\section{The Linear Boltzmann Transport model}
\label{LBT}

LBT model was developed to study jet propagation during the phase of quark-gluon plasma (QGP) in high-energy heavy-ion collisions with an emphasis on the jet-induced medium response in terms of the propagation of thermal recoil parton amid the evolving bulk medium as described by a relativistic hydrodynamic model. The model has been revamped recently with the implementation of the complete set of the elastic $2\to 2$ scattering processes \cite{He:2015pra}. Inelastic processes of medium induced $2\rightarrow 2+n$ multiple gluon radiation processes have also been implemented more consistently in the latest version of the LBT model. The LBT model is designed to describe not only parton energy loss and suppression of leading partons but also jet-induced medium excitation. It has been used to describe experimental data on suppression of single inclusive light and heavy flavor hadrons \cite{Cao:2016gvr,Cao:2017hhk}, single inclusive jets \cite{lbt-singlejet}, and $\gamma$-hadron \cite{Chen:2017zte} and $\gamma$-jet correlations \cite{Wang:2013cia}.

Parton transport for both jet shower and thermal recoil partons in the quark-gluon plasma are described by the linear Boltzmann equations,
\begin{eqnarray}
p_a\cdot\partial f_a&=&\int \prod_{i=b,c,d}\frac{d^3p_i}{2E_i(2\pi)^3} \frac{\gamma_b}{2}(f_cf_d-f_af_b)|{\cal M}_{ab\rightarrow cd}|^2
\nn && \hspace{-0.5in}\times
S_2(\hat s,\hat t,\hat u)(2\pi)^4\delta^4(p_a\!+\!p_b\!-\!p_c\!-\!p_d)+ {\rm inelastic},
\label{bteq}
\end{eqnarray}
where $\gamma_b$ is the color-spin degeneracy for parton $b$, $f_i=1/(e^{p_i\cdot u/T}\pm1)$ $(i=b,d)$ are thermal parton phase-space distributions in the QGP medium with local temperature $T$ and fluid velocity $u=(1, \vec{v})/\sqrt{1-\vec{v}^2}$, $f_i=(2\pi)^3\delta^3(\vec{p}-\vec{p_i})\delta^3(\vec{x}-\vec{x_i}-\vec{v_i}t)$ $(i=a,c)$ are the phase-space density for jet shower partons before and after scattering and medium recoil partons.  The collinear divergency  in the leading-order (LO) elastic scattering amplitude $|{\cal M}_{ab\rightarrow cd}|^2$ \cite{Eichten:1984eu} is regulated by a factor \cite{Auvinen:2009qm},
\begin{equation}
S_2(\hat s, \hat t, \hat u) = \theta(\hat s\ge 2\mu_{D}^2)\theta(-\hat s+\mu_{D}^2\le \hat t\le -\mu_{D}^2),
\end{equation}
 where $\hat s$, $\hat t$, and $\hat u$ are Mandelstam variables, and $\mu_{D}^2 = 3g^2 T^2/2$ is the Debye screen mass with 3 quark flavors. The cross section of corresponding elastic collision is $d\sigma_{ab\rightarrow cd}/d\hat t=|{\cal M}_{ab\rightarrow cd}|^2/16\pi \hat s^2$. The effect of quantum statistics in the final state and detailed balance are neglected in the current implementation of the Boltzmann transport model. The effective strong coupling constant $\alpha_s=g^{2}/4\pi$ is fixed and will be fitted to experimental data.

In the above linear Boltzmann transport equation, the inelastic processes include only induced gluon radiation accompanying elastic scattering in the current version of LBT. The radiative gluon spectrum is simulated according to the high-twist approach \cite{Guo:2000nz,Wang:2001ifa,Zhang:2003wk,Zhang:2004qm},
\ba \la{induced}
\frac{dN_g^{a}}{dzdk_\perp^2d\tau}=\frac{6\alpha_{\rm s}P_a(z)k_\perp^4}{\pi (k_\perp^2+z^2m^2)^4} \frac{p\cdot u}{p_0}\hat{q}_{a} (x)\sin^2\frac{\tau-\tau_i}{2\tau_f},
\ea
where $m$ is the mass of the propagating parton $a$, $z$  and $k_\perp$ are the energy fraction and transverse momentum of the radiated gluon, $P_a(z)$ the splitting function,  $\tau_f=2p_0z(1-z)/(k_\perp^2+z^2m^2)$ the gluon formation time and $\tau_i$ is the time of the last gluon emission.
The jet transport parameter,
\begin{equation}
\hat{q}_{a}(x)=\sum_{bcd}\rho_{b}(x)\int d\hat t q_\perp^2 \frac{d\sigma_{ab\rightarrow cd}}{d\hat t},
\end{equation}
is defined as the transverse momentum transfer squared per mean-free-path in the local comoving frame, where $\rho_{b}(x)$ is the parton density (including the degeneracy). The Debye screen mass $\mu_D$ is used as an infrared cut-off for the energy of the radiated gluons.

The probability of elastic and inelastic scattering in each time step are implemented together to ensure unitarity in LBT. The probability for an elastic scattering in a time step  $\Delta \tau $ during the propagation of parton $a$ is
\begin{equation}
P^a_{\rm el}=1-\text{exp}[- \Delta\tau \Gamma_a^{\rm el}(x)],
\end{equation}
where
\begin{equation}
\Gamma_a^{\rm el}\equiv \frac{p\cdot u}{p_0}\sum_{bcd} \rho_b(x)\sigma_{ab\rightarrow cd}
\end{equation}
is the elastic scattering rate.  The probability for inelastic process is
\begin{equation}
P^a_\mathrm{inel}=1-\exp[-\Delta\tau \Gamma_a^{\rm inel}(x)],
\end{equation}
where
\begin{equation}
\Gamma_a^{\rm inel}=\frac{1}{1+\delta_g^a}\int dz dk_\perp^2 \frac{dN^a_g}{dzdk_\perp^2d\tau}
\end{equation}
is the gluon radiation rate. The total scattering probability,
\begin{equation}
P^a_\mathrm{tot}=P^a_\mathrm{el}(1-P^a_\mathrm{inel}) +P^a_\mathrm{inel},
\end{equation}
can be separated  into the probability for pure elastic scattering (first term) and the probability for inelastic scattering with at least one gluon radiation (the second term). Multiple gluon radiation is simulated by a Poisson distribution with the mean $\langle N^a_g \rangle=\Delta\tau\Gamma_a^{\rm inel}$. The scattering channel, final flavor, energy and momentum of the final scattering partons, recoil partons and radiated gluons are sampled according to the elastic scattering amplitudes and the radiative gluon spectra, respectively. Global energy and momentum conservation is ensured in each scattering with multiple radiated gluons.

In LBT, all final partons after each scattering, including jet shower partons, recoil medium partons and radiated gluons,  will go through further scattering in the medium. To account for the back reaction in the Boltzmann transport, initial thermal parton $b$ in each scattering are tracked as ``negative'' partons and they also propagate in the medium according to the Boltzmann equation. These ``negative" partons  are part of the jet-induced medium excitation and manifest as the diffusion wake behind the propagating jet shower partons \cite{Wang:2013cia,Li:2010ts,He:2015pra}. The energy and momentum of these ``negative" partons will be subtracted from all final observables.

A hydrodynamic model is used to provide spatial and time information on the local temperature and fluid velocity of the bulk QGP medium which evolves independently of the jet propagation. In the linear approximation ($\delta f\ll f$), we neglect interaction among jet shower and recoil partons and consider only interaction of jet shower and recoil partons with thermal medium partons. This assumption will break down when the jet-induced medium excitation becomes appreciable relative to the local thermal parton density.   To extend LBT beyond this region of applicability, a coupled LBT and hydrodynamic (CoLBT-hydro) model \cite{Chen:2017zte} has been developed in which jet transport is coupled to hydrodynamic evolution of the bulk medium in real time through a source term in the hydrodynamic equations from the energy and momentum deposited by propagating jet shower partons. This coupled approach assumes complete local equilibration of soft partons from LBT and most suitable for the study of hadron spectra from jet-induced medium excitation.

In the current version of LBT model, the parton recombination model \cite{Han:2016uhh} developed by the Texas A \& M University group within the JET Collaboration is employed for hadronization of all jet shower and thermal recoil partons. The model has been employed successfully to study  $\gamma$-jet modification, light and heavy flavor hadron suppression in heavy-ion collisions \cite{Wang:2013cia,Cao:2016gvr,Cao:2017hhk}. In this paper, we will only use the partonic information for jet reconstruction and study of jet profiles.

\section{$\gamma$-jet asymmetry}
\label{asym}

To study $\gamma$-jet correlations, we use PYTHIA 8~\cite{Sjostrand:2006za,Sjostrand:2007gs} to generate the initial jet shower partons in $\gamma$-jet events in $p+p$ collisions. We generate $\gamma$-jet events with a minimum transverse momentum transfer of the hard processes that is half of the transverse momentum of the triggered photons. These events also include bremsstrahlung photons from QCD processes. In principle, the photon bremsstrahlung processes can also be modified by the final state interaction and photons can also be produced through jet-medium interaction \cite{Zakharov:2004bi,Turbide:2005fk}. However, for the large values of transverse momentum of  photons we consider in this study, the modification is negligible as in the case of bremsstrahlung production of heavy quark pairs \cite{Cao:2015kvb}.

  The background hydrodynamic profile for the bulk QGP medium is provided by simulations from CLVisc (3+1)D viscous hydrodynamics model ~\cite{Pang:2012he,Pang:2014ipa,Pang:2018zzo} with initial conditions from A Multi-Phase Transport (AMPT) model \cite{Lin:2004en}. The initial energy-momentum density is normalized at the initial time $\tau_0=0.4$ fm/$c$ so that the final hadron spectra from the hydrodynamic simulations with freeze-out temperature $T_{\rm f}=137$ MeV can reproduce experimental data on the final charged hadron rapidity and transverse momentum distributions \cite{Pang:2012he,Pang:2018zzo,Pang:2015zrq,Pang:2014pxa}.
The centralities of heavy-ion collisions are selected according to the fractional event distribution in the initial parton multiplicity in the central rapidity region. The spatial distribution of the initial production points of $\gamma$-jets is sampled according to the initial hard parton distribution in the transverse plane as given in the same AMPT simulations. Each parton is assigned an initial formation time $\tau_{f0}={\rm Max}(\tau_0, 2E/p_T^2)$ and is allowed to interact with medium partons according to the Boltzmann transport only after this initial formation time. Transport of partons in the bulk medium is simulated until the thermal parton density vanishes after the QCD phase transition during the hydrodynamic evolution and partons free-stream afterwards.  We use  FASTJET \cite{Cacciari:2011ma} to reconstruct jets from the final partons after parton transport within LBT. $\gamma$-jet pairs are selected within the same kinematic cuts as imposed in the experimental measurements that we will compare to. In CMS experimental data, kinetic cuts $|\eta_\gamma|<1.44$, $|\eta_{\rm jet}|<1.6$ and $|\phi_\gamma-\phi_{\rm jet}|>(7/8)\pi$ are imposed for all asymmetry studies.
\begin{figure}[!htb]
\centering
\vspace{-0.5in}
\includegraphics[width=6.7cm,bb=15 15 500 500]{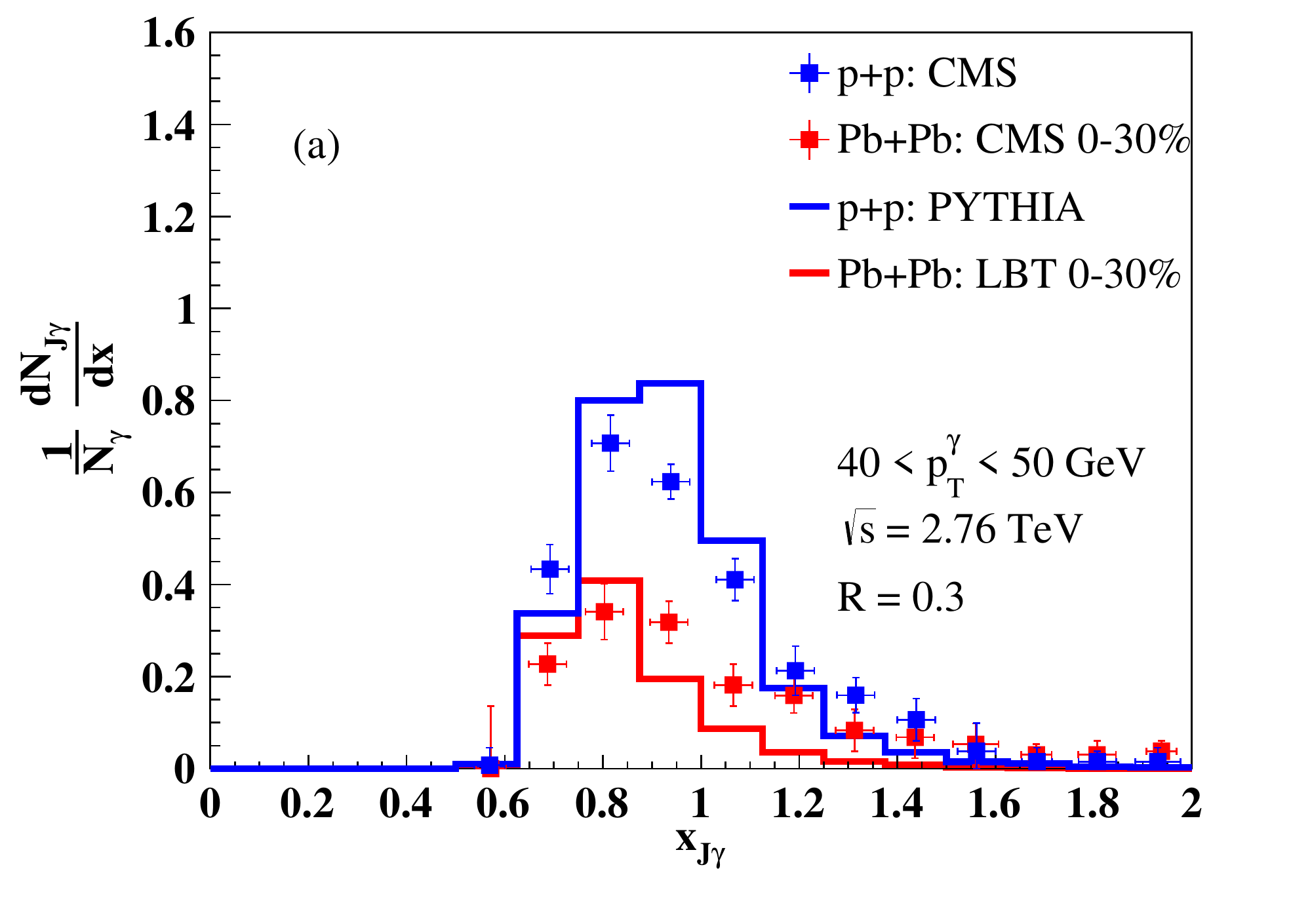} \\
\vspace{-0.69in}
\includegraphics[width=6.7cm,bb=15 15 500 500]{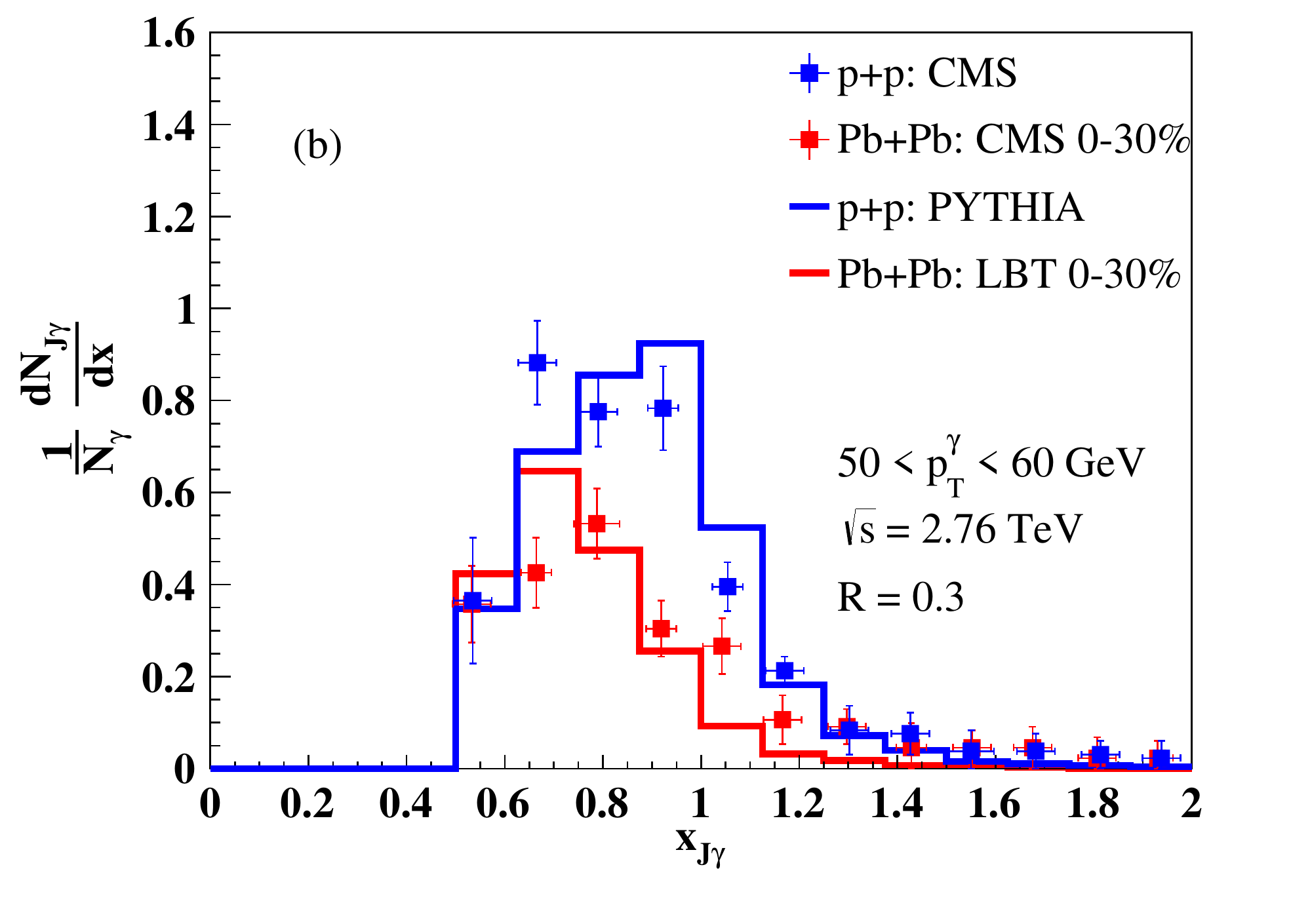} \\
\vspace{-0.69in}
\includegraphics[width=6.7cm,bb=15 15 500 500]{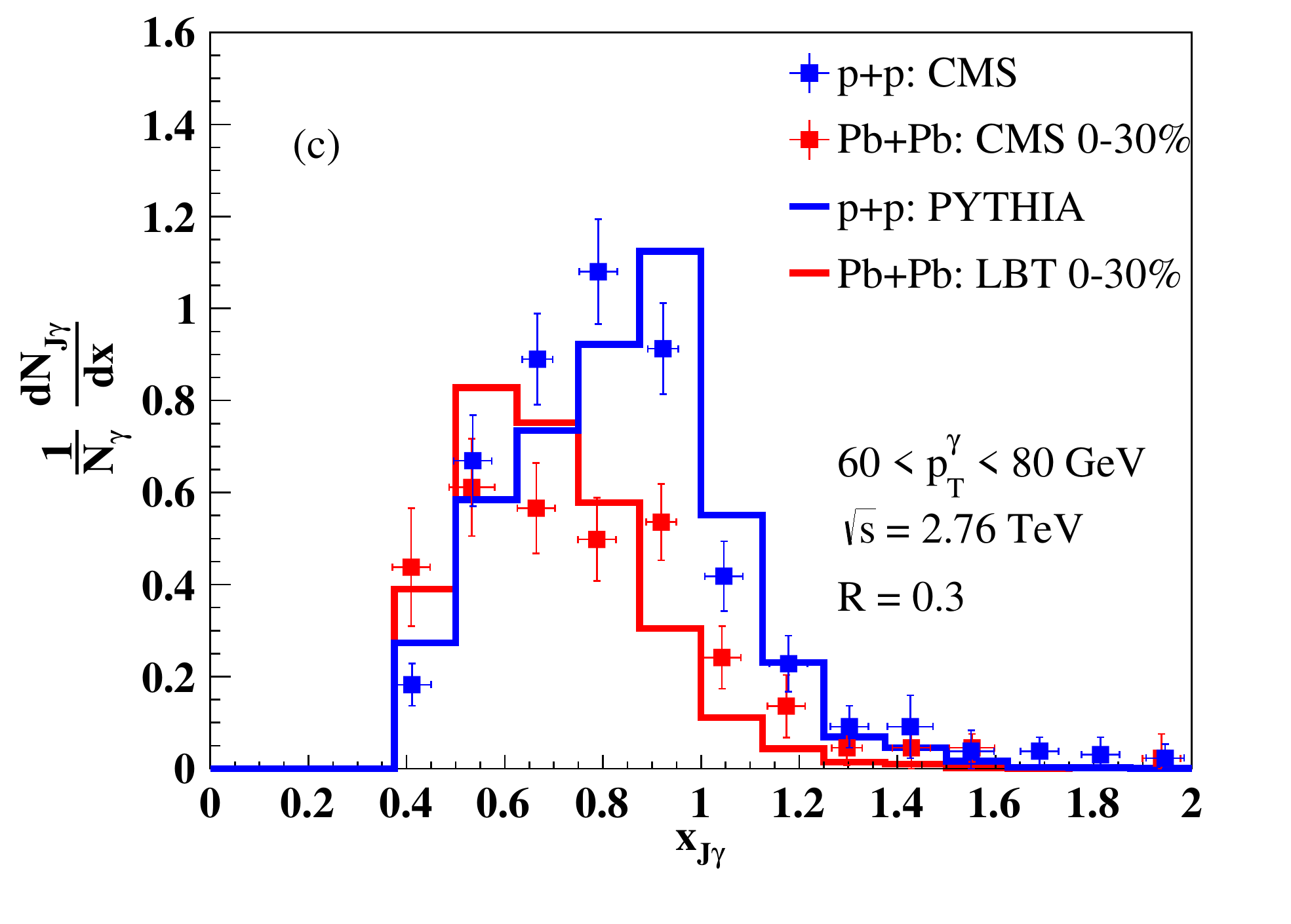} \\
\vspace{-0.69in}
\includegraphics[width=6.7cm,bb=15 15 500 500]{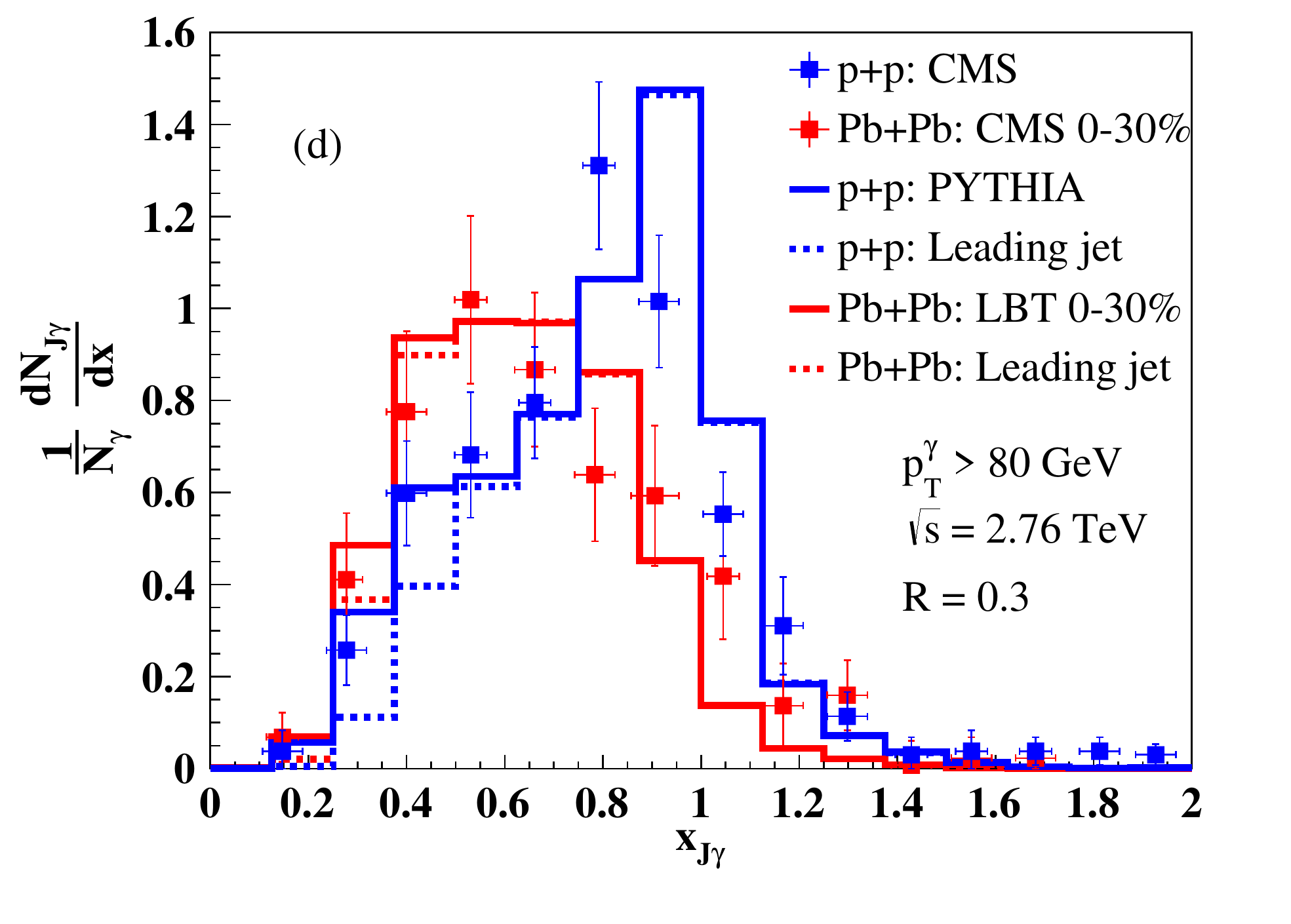}
\vspace{-0.1in}
\caption{(Color online) $\gamma$-jet asymmetry $x_{J\gamma}=\pt^\text{jet}/\pt^\gamma$ distribution in central (0--30\%) Pb+Pb collisions (red) and p+p collision (blue) for different values of $p_T^\gamma$ at $\sqrt{s}=2.76$ TeV from LBT simulations as compared to the CMS experimental data~\cite{Chatrchyan:2012gt}. Dashed lines in the bottom panel are the asymmetry distributions for leading jets only.}
\label{asymmetry}
\end{figure}

In order to compare jet measurements in Pb+Pb to p+p collisions in CMS experiment, the jet energy in p+p events is smeared with a Gaussian convolution to match the jet energy resolution in each of the Pb+Pb centrality classes in which the comparison is made. The parameters of the Gaussian smearing are centrality dependent \cite{Chatrchyan:2012gt}. In order to compare to the CMS experimental data, we also convolute our LBT results with the same Gaussian smearing. All LBT results in both Pb+Pb and p+p collisions are smeared with the same jet energy resolution in each centrality class of Pb+Pb collisions as in the CMS data  that we compare with. In principle, one should also subtract the contribution from underlying event to the jet energy as carried out in experimental analyses which vary from one experiment to another.   In the calculation of jet transverse energy in FASTJET, we assume there is a complete subtraction of the underlying event background that are not correlated to the triggered photon. The subtraction of the combinatory background from events mixed from $\gamma$-triggered and minimal biased events by CMS \cite{Chatrchyan:2012gt} is close to such a complete subtraction. Using the underlying event subtraction scheme as in ATLAS experiment \cite{Aad:2012vca}, the underlying event contribution in the LBT simulations is also small and the underlying event subtraction only reduce the energy of final reconstructed jet slightly \cite{lbt-singlejet}.

\begin{figure}[!htb]
\centering
\vspace{-0.5in}
\includegraphics[width=6.7cm,bb=15 15 500 500]{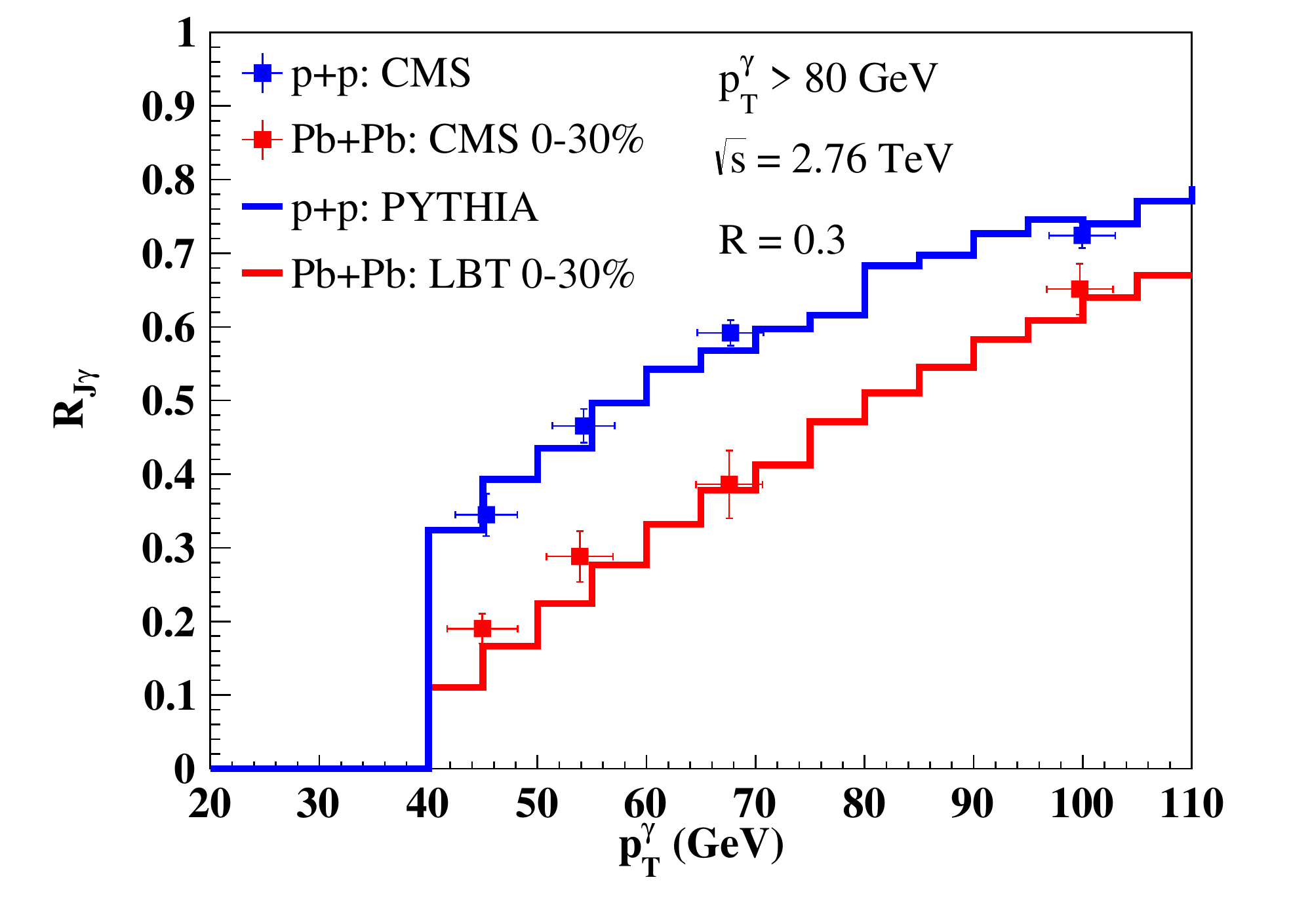} \\
\caption{(Color online) Fraction of photons associated with a jet with $p_T^{\rm jet}\ge 30$ GeV/$c$ as a function of $p_T^\gamma$ in central (0--30\%) Pb+Pb (red) and p+p collisions (blue) at $\sqrt{s}=2.76$ TeV from LBT simulations as compared to the CMS experimental data~\cite{Chatrchyan:2012gt}.}
\label{Rjg}
\end{figure}

Shown in Fig.~\ref{asymmetry} are results on $\gamma$-jet asymmetry $x_{J\gamma}=p_T^{\rm jet}/p_T^\gamma$ distributions from LBT simulations of $p+p$ and
0-30\% central Pb+Pb collisions at $\sqrt{s}=2.76$ TeV for different values of $p_T^\gamma$ as compared to CMS experimental data~\cite{Chatrchyan:2012gt}.
The jet cone-size is chosen as $R=0.3$ and the lower threshold of transverse momentum for reconstructed jets is $p_T^{\rm jet}>30$ GeV/$c$.   The LBT results with a fixed value of  $\alpha_{\rm s}=0.19 $ agree reasonably well with the experimental data. We should note that this value of $\alpha_{\rm s}$ is only the effective strong coupling constant in the jet-medium interaction within the LBT model which employs the perturbative estimate of Debye screening mass as a cut-off in the transverse momentum transfer in the elastic scattering between jet-shower/recoil and medium partons. An alternative cut-off and inclusion of non-perturbative processes such as magnetic screening \cite{Buzzatti:2011vt} can change the jet-medium interaction strength and lead to a larger effective value of $\alpha_{\rm s}$.

The asymmetry distributions are the absolute yields of jets above the threshold that are associated with the triggered $\gamma$. Therefore, one observes not only the shift of the peak of the distributions toward smaller asymmetry values in Pb+Pb relative to p+p collisions due to jet energy loss but also the reduction of the total jet yields above the cut-off $p_T^{\rm jet}>30$ GeV/$c$. Shown in Fig.~\ref{Rjg} are the fractions of isolated photons that have at least one associated jet above the cut-off $p_T^{\rm jet}>30$ GeV/$c$ as a function of $p_T^\gamma$ in both p+p and 0-30\% central Pb+Pb collisions. The fraction increases with $p_T^\gamma$ in both p+p and Pb+Pb collisions. It is however suppressed in central Pb+Pb collisions relative to p+p due to jet quenching.

\begin{figure}[!htb]
\centering
\vspace{-0.5in}
\includegraphics[width=6.7cm,bb=15 15 500 500]{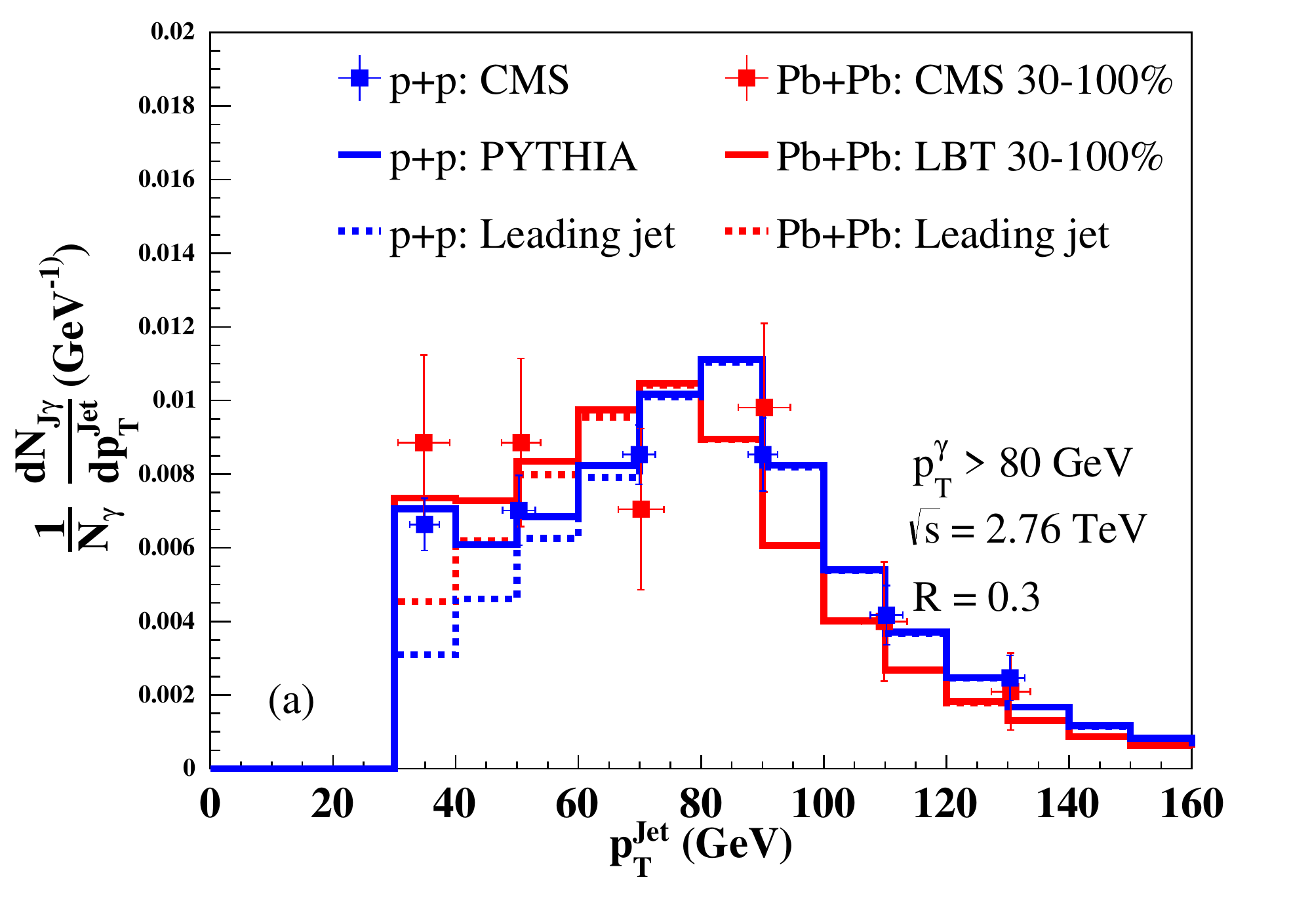}\\
\vspace{-0.69in}
\includegraphics[width=6.7cm,bb=15 15 500 500]{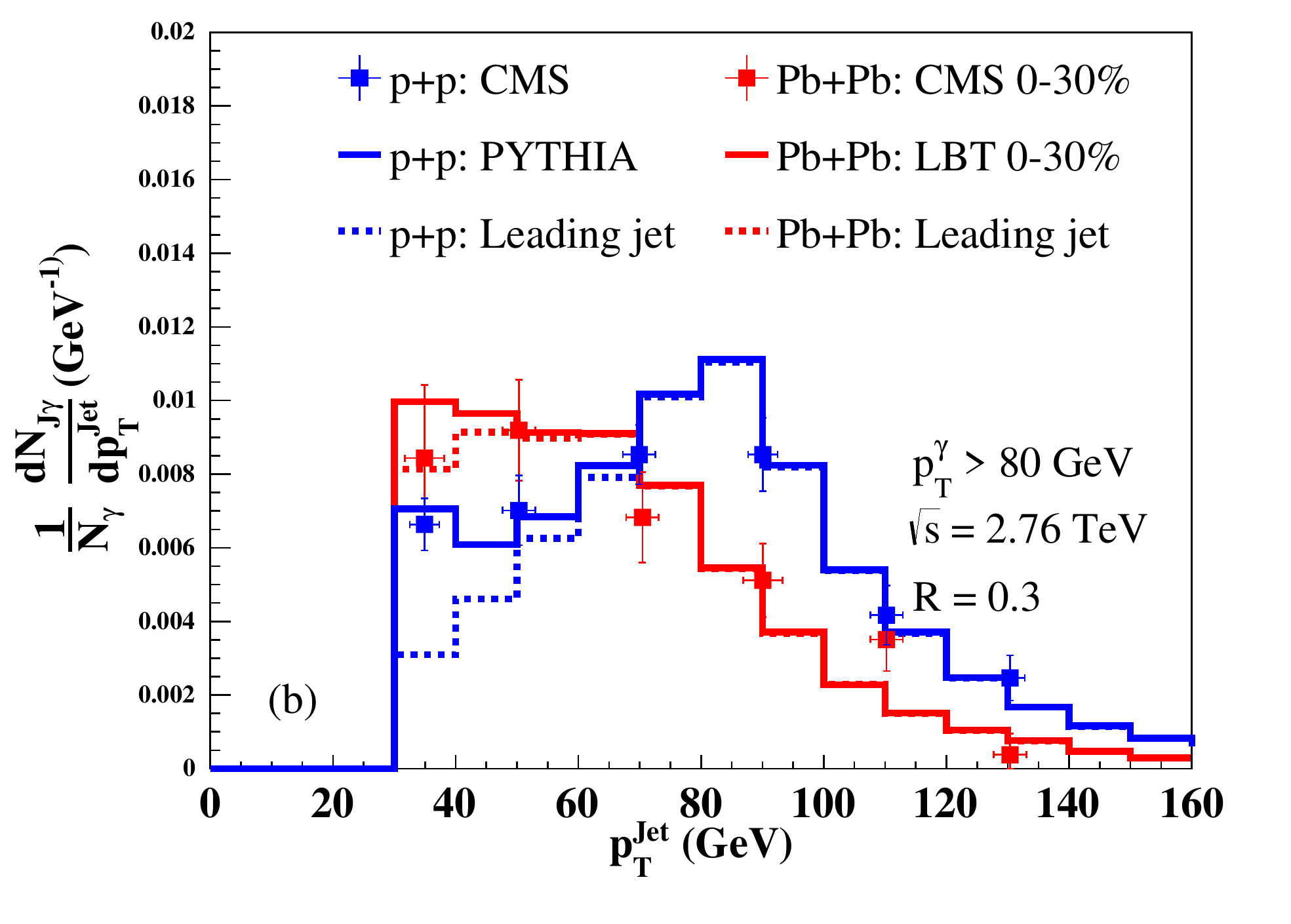} 
\caption{(Color online) (a) Transverse momentum distribution of $\gamma$-jet in peripheral (30--100\%) and (b) central (0--30\%) Pb+Pb (red) and p+p collisions (blue) at $\sqrt{s}=2.76$ TeV from LBT simulations as compared to the CMS experimental data ~\cite{Chatrchyan:2012gt}. Dashed lines are the transverse momentum distributions for leading jets only.}
\label{ptjet}
\end{figure}

The asymmetry distributions from both CMS experiment and LBT results (solid lines) shown in Fig.~\ref{asymmetry}  are the absolute jet yields associated with the tagged photon. The yields include not only the leading jet but also sub-leading jets if multiple jets are produced in association with the tagged photon. These multiple jets are produced through higher-order QCD processes and are kinematically possible when $p_T^\gamma$ is much larger than the cut-off of jet transverse momentum. We also show the LBT results for associated leading jet as dashed lines in the bottom panel of Fig.~\ref{asymmetry} which are smaller than the inclusive associated jet yields at small value of asymmetry $x_{J\gamma}$. The difference is mainly caused by sub-leading jets associated with the direct photon production.

To study the modification of associated jet spectra and illustrate the contribution of multiple jets, we show in Fig.~\ref{ptjet} the distribution of associated jet yields as a function of $p_T^{\rm jet}$ for fixed $p_T^\gamma \ge 80$ GeV/$c$ in both p+p and Pb+Pb collisions with two different centralities at $\sqrt{s}=2.76$ TeV. The LBT results compare fairly well with the experimental data from CMS \cite{Chatrchyan:2012gt}. We also show the LBT results for associated leading jet as dashed lines which deviate from the inclusive associated jet yields at small value of $p_T^{\rm jet}$. The difference between inclusive and leading jet yield is the contribution from mainly secondary jets associated with the direct photon production. As the LBT results indicate, these secondary jets are produced at lower $p_T^{\rm jet}$. Because of jet energy loss in medium, the peak of the leading jet distribution is shifted to a smaller value of $p_T^{\rm jet}$ in Pb+Pb collisions relative to that in p+p collisions. The contribution from the secondary jets to the inclusive jet yield above the cut-off $p_T^{\rm jet}>30$ GeV/$c$ is also suppressed in Pb+Pb collisions due to jet energy loss. In 0-30\% central Pb+Pb collisions, the jet energy loss shifts the peak of the inclusive jet distribution to a smaller value of $p_T^{\rm jet}$ very close to the cut-off and the peak is further smeared by the contribution from the secondary jets. The peak structure of the associated leading jet distribution is much pronounced in non-central Pb+Pb collisions when $p_T^\gamma$ is much larger than the cut-off of $p_T^{\rm jet}$ of the reconstructed jets. 

\begin{figure}[!htb]
\centering
\vspace{-0.5in}
\includegraphics[width=6.7cm,bb=15 15 500 500]{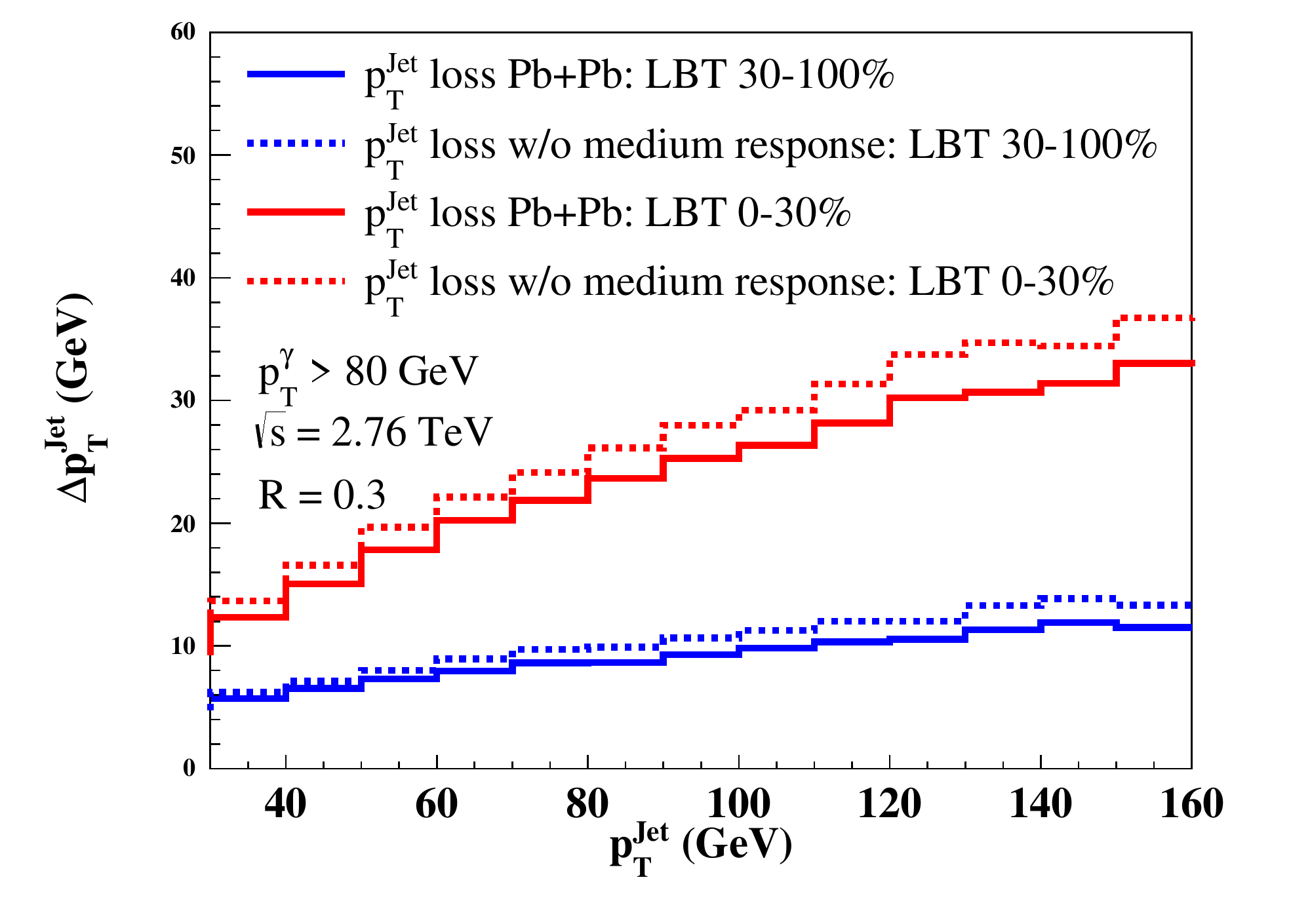} \\
\caption{(Color online) 
Average transverse momentum loss of the leading $\gamma$-jet in two centrality classes of Pb+Pb collisions at $\sqrt{s}=2.76$ TeV calculated within LBT as a function of the initial jet transverse momentum with (solid) and without (dashed) contributions from recoil and ``negative" partons in the jet reconstruction.}
\label{ptloss}
\end{figure}

The associated jet spectra for a fixed value of $p_T^\gamma$ shown in Fig.~\ref{ptjet}  should be a better direct measurement of the total jet energy loss through the shift of the spectra in Pb+Pb relative to p+p collisions. To illustrate this, we show in Fig.~\ref{ptloss} the calculated average transverse momentum loss of the leading jet in $\gamma$-jet events in two centrality classes of Pb+Pb collisions at $\sqrt{s}=2.76$ TeV as a function of the transverse momentum of the initial leading jet before their transport and propagation in the bulk medium. The solid (dashed) lines are the transverse momentum loss for leading $\gamma$-jets that (do not) include recoil and ``negative" partons from medium response in the jet reconstruction. We observe that the jet transverse momentum loss due to jet quenching increases with the initial jet transverse momentum and the dependence is slightly less than a linear increase. This transverse momentum dependence is a combined effect of energy dependence of the jet energy loss for a given jet flavor (quark or gluon) and the transverse momentum dependence of initial jet flavor composition. The initial fraction of quark jets increases with transverse momentum in the $\gamma$-jet production processes and the energy loss of a gluon jet is found to be about 1.5 times bigger than that of a quark for a jet cone-size $R=0.3$ in this range of $p_T^{\rm jet}$ \cite{lbt-singlejet}.

\begin{figure}[!htb]
\centering
\vspace{-0.5in}
\includegraphics[width=6.7cm,bb=15 15 500 500]{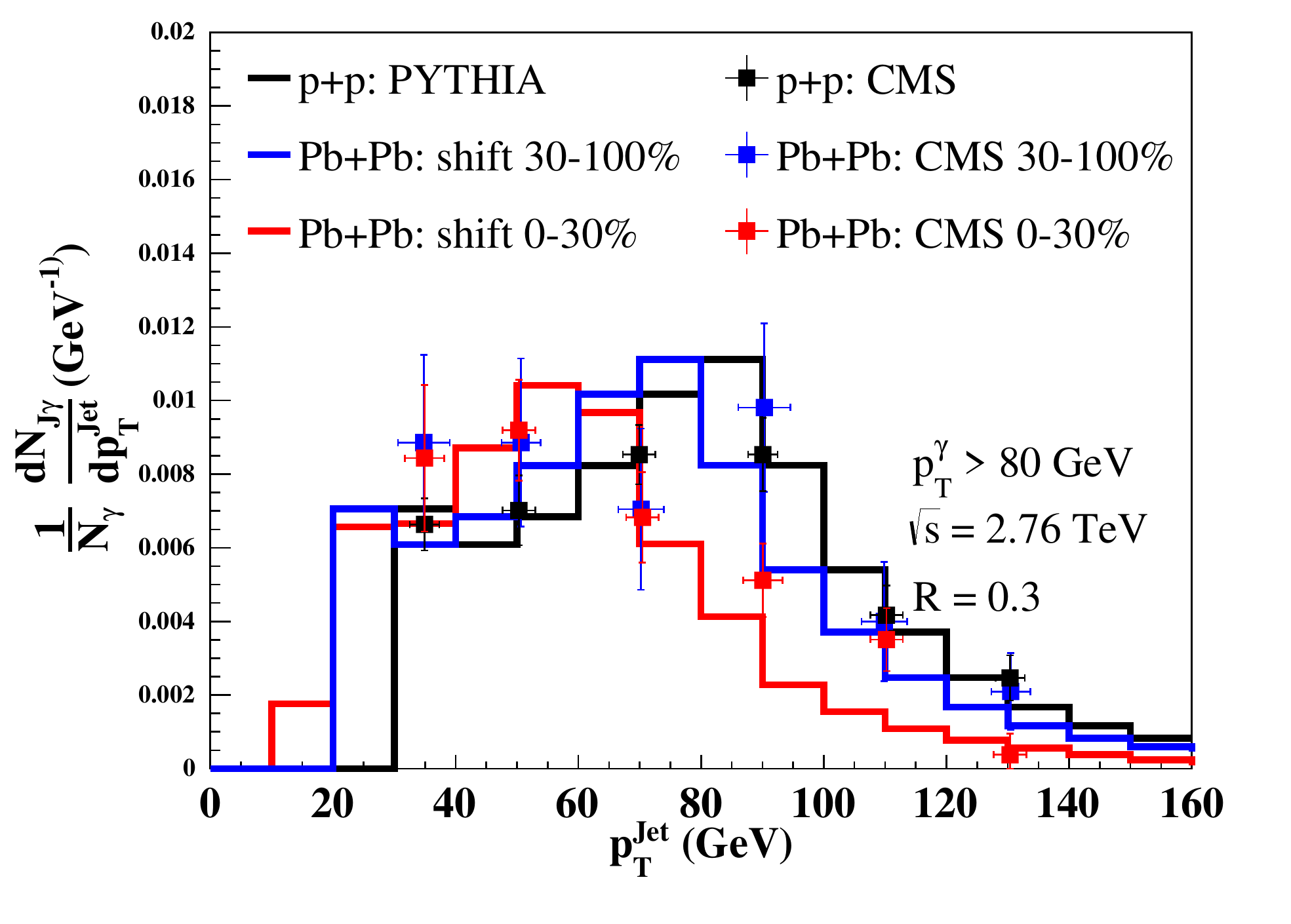} \\
\caption{(Color online) 
Associated  $\gamma$-jet spectrum as a function of $p_T^{\rm jet}$ for fixed $p_T^\gamma \ge 80$ GeV/$c$ in p+p collisions  (black) and spectra via shifting the p+p spectrum with the average transverse momentum loss in two centrality classes of Pb+Pb collisions (blue and red) at $\sqrt{s}=2.76$ TeV compared with the CMS experimental data \cite{Chatrchyan:2012gt}. The p+p spectrum is smeared according to the CMS parameters for jet energy resolution in 0-30\% Pb+Pb collisions.}
\label{ptshift}
\end{figure}

One can estimate the medium modification of the associated $\gamma$-jet spectra in Pb+Pb collisions by shifting the spectrum in p+p collisions with the average transverse momentum loss as calculated in LBT. The modified $\gamma$-jet spectra via shifting the p+p spectrum compare quite well with the calculated spectra within LBT and CMS data as shown in Fig.~\ref{ptshift} for two (0-30\%, 30-100\%) centrality classes of Pb+Pb collisions at $\sqrt{s}=2.76$ TeV.  The p+p spectrum used for obtained modified spectra via transverse momentum shift is smeared according to the CMS parameters for jet energy resolution in 0-30\% Pb+Pb collisions.  For more precision estimate of the spectra via shifting p+p jet spectra one should take into account the fluctuation in jet energy loss due to variation of the jet propagation path length. Such a method can in principle serve as a more direct extraction of jet energy loss from experimentally measured jet spectra. 
 

\section{$\gamma$-jet azimuthal angle correlation}
\label{azym}

\begin{figure}[!htb]
\centering
\vspace{-0.5in}
\includegraphics[width=6.7cm,bb=15 15 500 500]{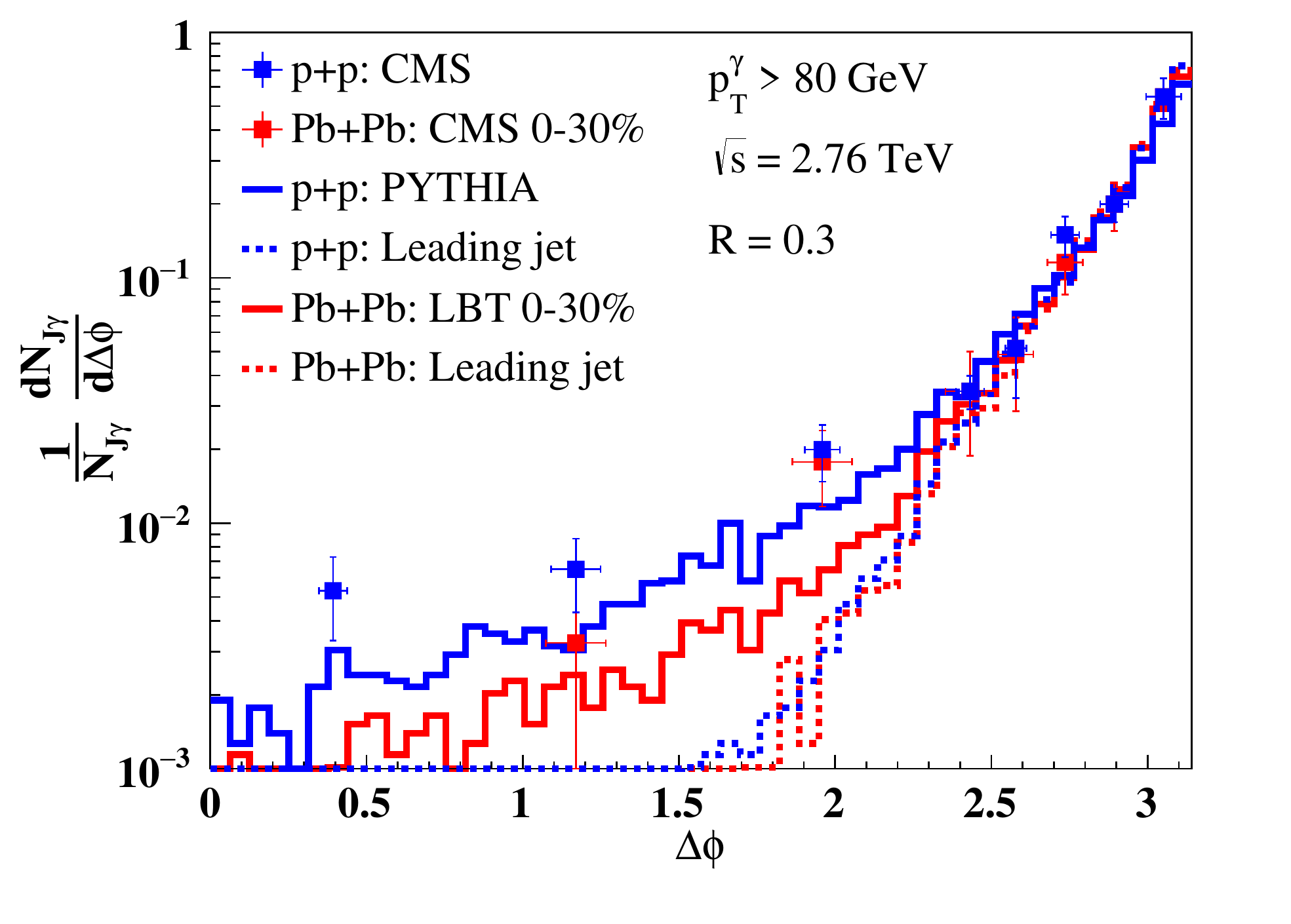}
\caption{(Color online) Angular distribution of $\gamma$-jet in central (0--30\%) Pb+Pb (red) and p+p collisions (blue) at $\sqrt{s}=2.76$ TeV from LBT simulations as compared to the CMS experimental data. Dashed lines are the angular correlations of leading jets in the $\gamma$-jet events.}
\label{dndphi}
\end{figure}

\begin{figure}[!htb]
\centering
\vspace{-0.5in}
\includegraphics[width=6.7cm,bb=15 15 500 500]{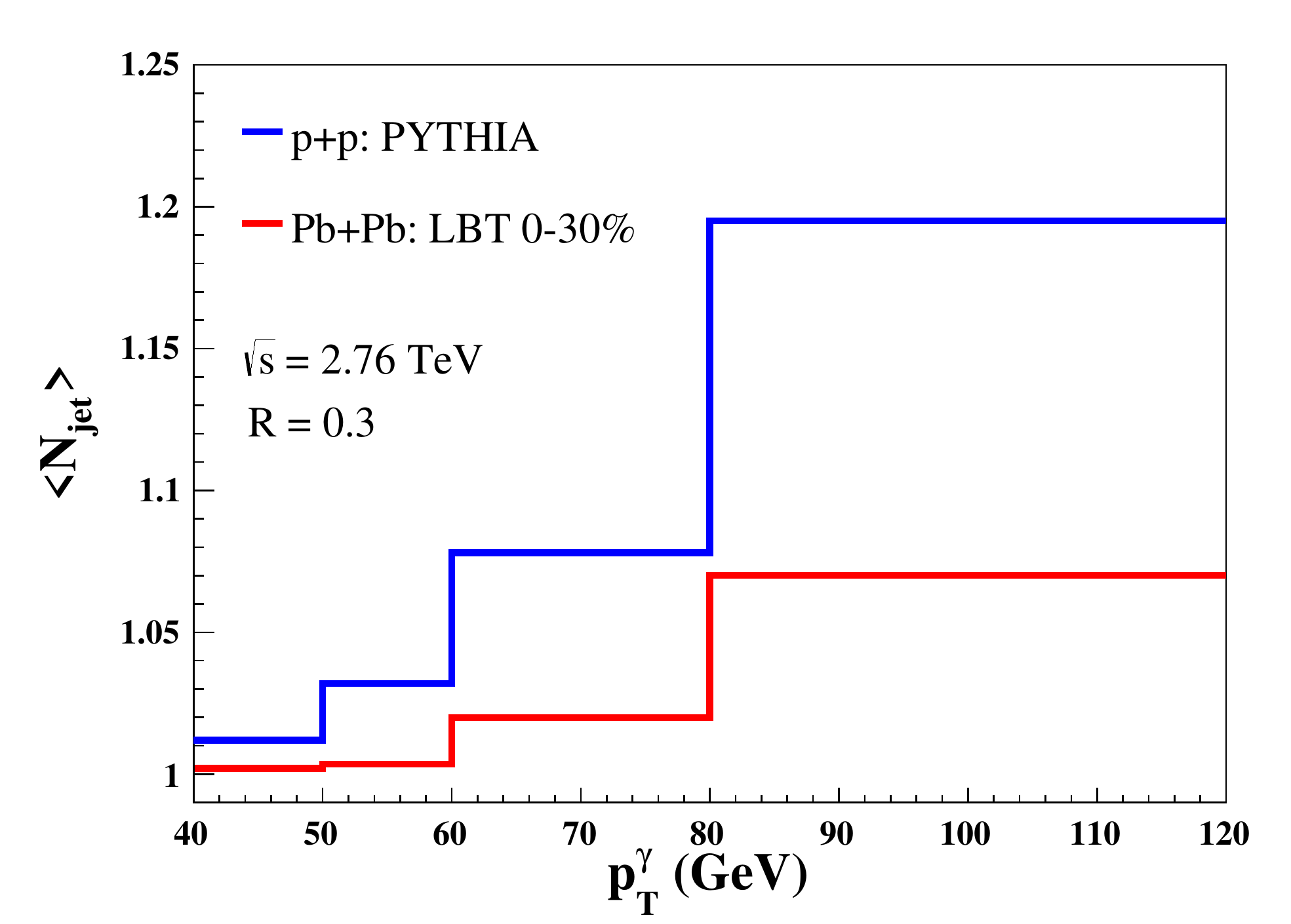}
\caption{(Color online)  Number of jets per $\gamma$-jet events as a function of transverse momentum of the triggered photon in p+p (blue) and central (0--30\%) Pb+Pb collisions (red).}
\label{njet}
\end{figure}

To further investigate the contribution of sub-leading jets in $\gamma$-jet correlation, we show in Fig.~\ref{dndphi} the $\gamma$-jet correlation in azimuthal angle $\Delta \phi_{\gamma J}=|\phi_\gamma -\phi_{\rm jet}|$ from LBT simulations and CMS experiment data for p+p and 0-30\% central Pb+Pb collisions at $\sqrt{s}=2.76$ TeV. Note that this a distribution normalized to $\gamma$-jet pairs for $p_T^\gamma>80$ GeV/$c$ and $p_T^{\rm jet}>30$ GeV/$c$. The azimuthal angle correlation for leading $\gamma$-jet as plotted in dashed lines are much narrower than the inclusive $\gamma$-jet correlation and the distribution for Pb+Pb collisions is almost indistinguishable from that in p+p collisions. This is consistent with the conclusion of Refs.~\cite{Mueller:2016gko,Chen:2016vem} that angular correlation for large values of $p_T^\gamma$ is dominated by the Sudakov form factor from soft initial state radiation and the effect of transverse momentum broadening due to multiple scattering in medium is negligible. The $\gamma$-jet correlation at large azimuthal angle $\Delta \phi_{\gamma J}$ is dominated by sub-leading jets from the next-to-leading processes of multiple jet production. Because of suppression of sub-leading jets due to jet energy loss, their contribution to the inclusive $\gamma$-jet correlation at large angle is also reduced. This leads to an apparent suppression of single inclusive $\gamma$-jet correlation at large angle in Pb+Pb relative to p+p collisions. The contribution of multiple jets to and their suppression in the $\gamma$-jet correlation in heavy-ion collisions will therefore make it difficult to use the angular correlation to study the structure of QGP through large angle scattering \cite{DEramo:2012uzl}. The contribution from multiple jets becomes more important for large values of $p_T^\gamma$ that is significantly larger than $p_T^{\rm jet}$. Shown in Fig.~\ref{njet} is the average number of jets per $\gamma$-trigger for $p_T^{\rm jet}>$ 30 GeV/$c$ as a function of $p_T^\gamma$ in p+p and 0-30\% central Pb+Pb collisions at $\sqrt{s}=2.76$ GeV. One can see that the average jet yield per $\gamma$ trigger increases with $p_T^\gamma$ and is suppressed in Pb+Pb relative p+p collisions due to jet energy loss.

\section{$\gamma$-jet profile and energy flow}
\label{flow}

\begin{figure}[!htb]
\centering
\vspace{-0.5in}
\includegraphics[width=6.7cm,bb=15 15 500 500]{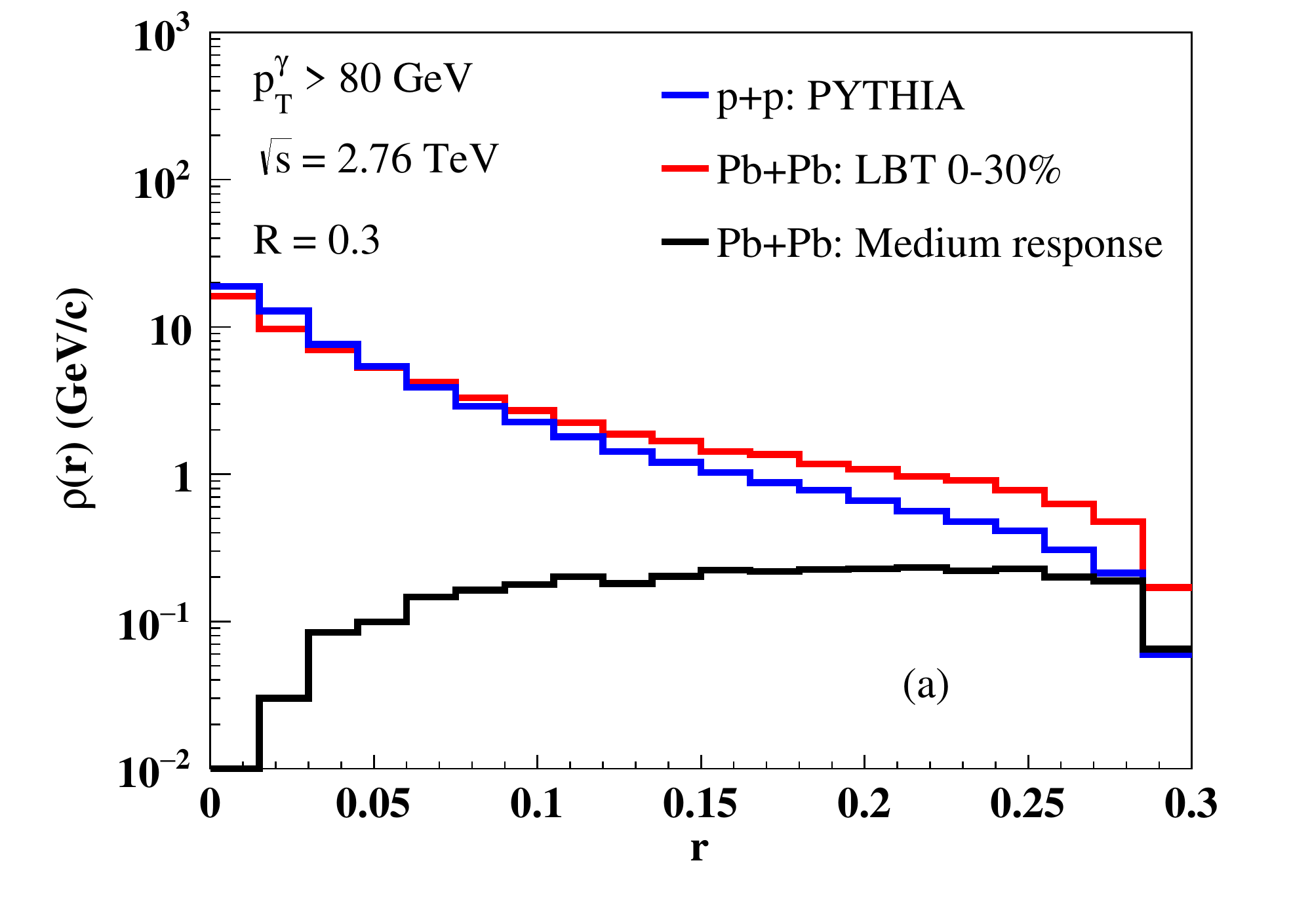} \\
\vspace{-0.69in}
\includegraphics[width=6.7cm,bb=15 15 500 500]{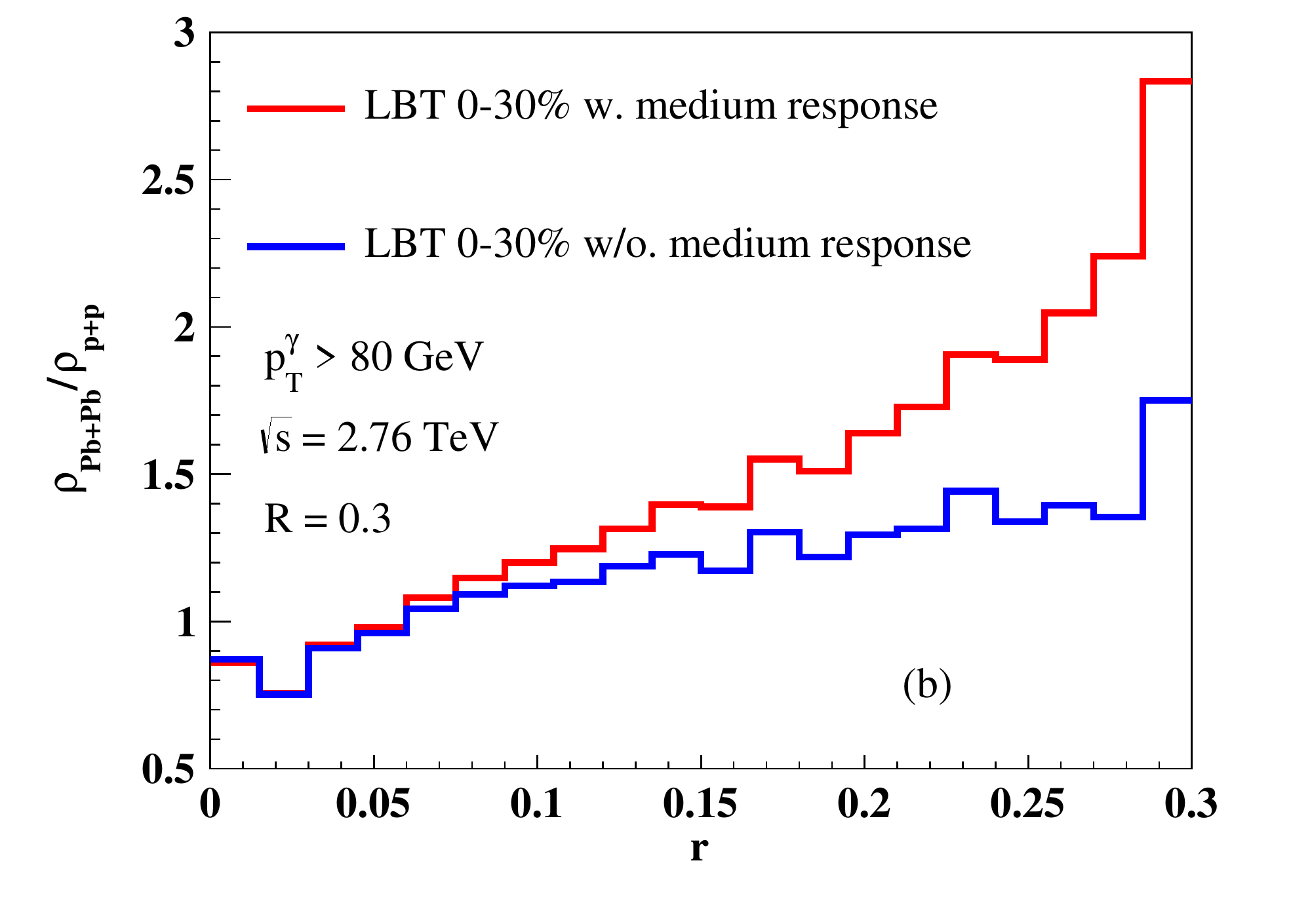}
\caption{(Color online) (a) The ``exclusive" transverse jet profile of $\gamma$-jets in central (0--30\%) Pb+Pb (red) and p+p collisions (blue) from LBT simulations and (b) their ratio with (red) and without (blue) contributions from jet-induced medium response at $\sqrt{s}=2.76$ TeV. The black line in (a) is the contribution to the jet transverse profile from jet-induced medium response (recoil and ``negative" partons). }
\label{profile1}
\end{figure}

To study how the energy lost by leading partons is transported through the medium, we investigate the modification of energy distribution inside the jet cone or the jet transverse profile, 
\begin{equation}
\rho(r)=\frac{1}{\Delta r}\frac{1}{N_\text{jet}}\sum_\text{jet}\frac{p_T^{\rm jet}(r-\Delta r/2,r+\Delta r/2)}{p_T^{\rm jet}(0,R)},
\label{eq:profile}
\end{equation}
defined as the distribution of summed transverse energy,
\begin{equation}
p_T^{\rm jet}(r-\Delta r/2,r+\Delta r/2)=\sum_{{\rm assoc}\in \Delta r} p_T^{\rm assoc},
\end{equation}
carried by the associated particles within a circular annulus with a width $\Delta r$ at radius $r=\sqrt{(\eta-\eta_{\rm jet})^2+(\eta-\eta_{\rm jet})^2}$ normalized to the total transverse energy within the jet cone $r\in R$, where $\eta_{\rm jet}$ and $\phi_{\rm jet}$ are the location of the center of the jet given by FASTJET. The definition of the jet profile can be slightly different depending on which particles are used to calculate the total transverse momentum within a circular annulus of the jet cone. We first consider the jet profile calculated from the associated partons from the list of particles within the reconstructed jet in FASTJET.  We refer to this jet profile as ``exclusive". Shown in Fig.~\ref{profile1} (upper panel) is the ``exclusive" transverse profile  of  $\gamma$-jets in p+p and central 0-30\% Pb+Pb collisions and their ratio  (lower panel) at $\sqrt{s}=2.76$ TeV.  One can see that jet-medium interaction has transported certain amount of energy toward the outer layer of the jet cone and leads to significant enhancement of jet transverse profile at large $r$ near the edge of jet cone. This enhancement is compensated by slight suppression of transverse energy at the very core ($r<0.03$) of the jet. In Fig.~\ref{profile1}(a), we also plot the contribution to the jet transverse profile from jet-induced medium response (recoil and ``negative partons") in Pb+Pb collisions, which becomes more signifiant toward the edge of the circular jet cone. The enhancement of the jet transverse profile in Pb+Pb over that in p+p collisions as seen in the ratio in Fig.~\ref{profile1}(b) is caused mainly by the medium response, without which the enhancement is significantly smaller as shown in Fig.~\ref{profile1}(b).  Jet quenching also changes the flavor composition of the final jet since gluon-initiated jets lose more energy than quark-initiated jets. This effect will lead to a larger fraction of quark-initiated jets which could narrow the jet profile since the quark jet profile is narrower than that of a gluon. This will compete with the broadening of the jet profile due to induced gluon radiation and jet-induced medium response.

\begin{figure}[!htb]
\centering
\vspace{-0.5in}
\includegraphics[width=6.7cm,bb=15 15 500 500]{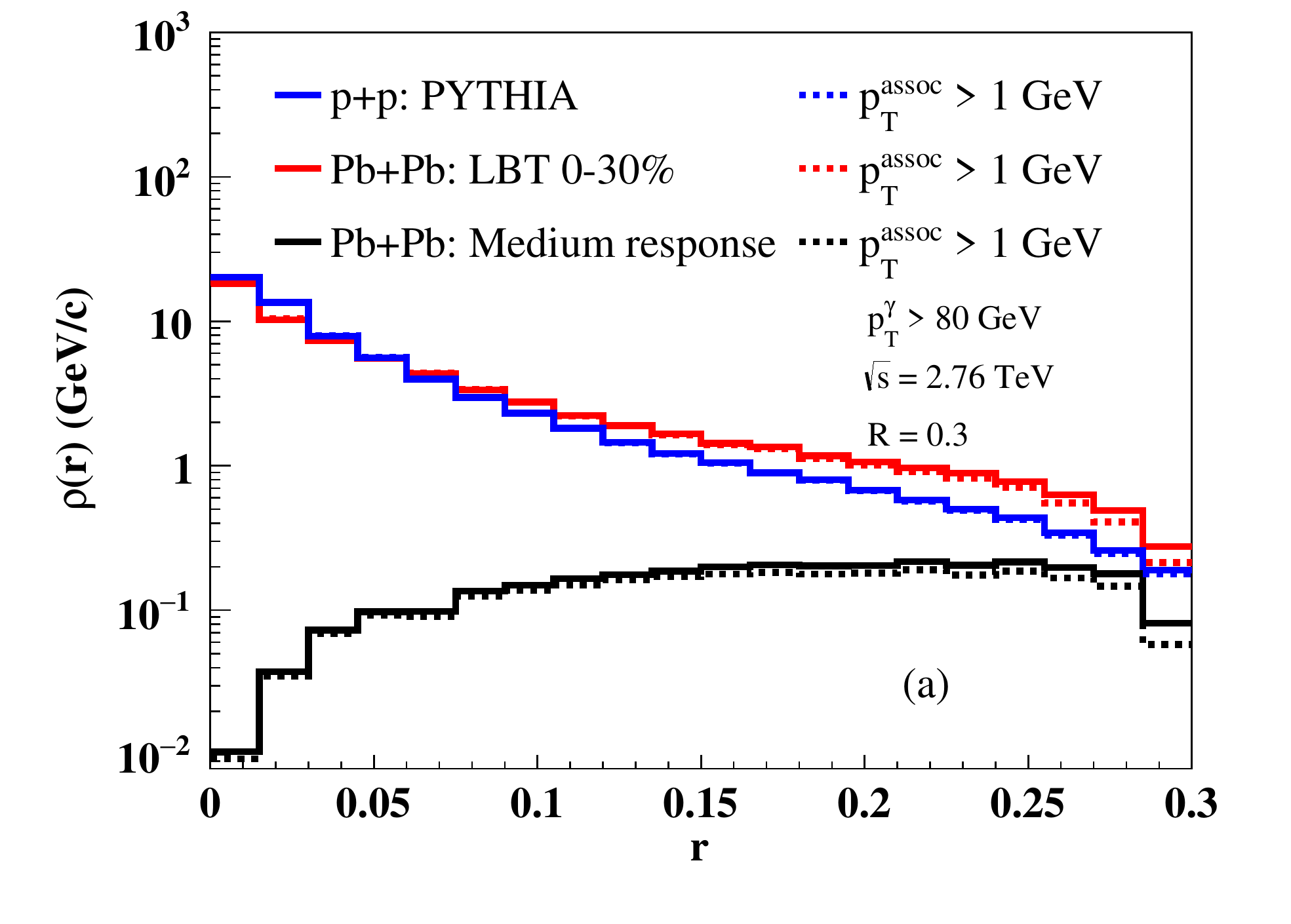} \\
\vspace{-0.69in}
\includegraphics[width=6.7cm,bb=15 15 500 500]{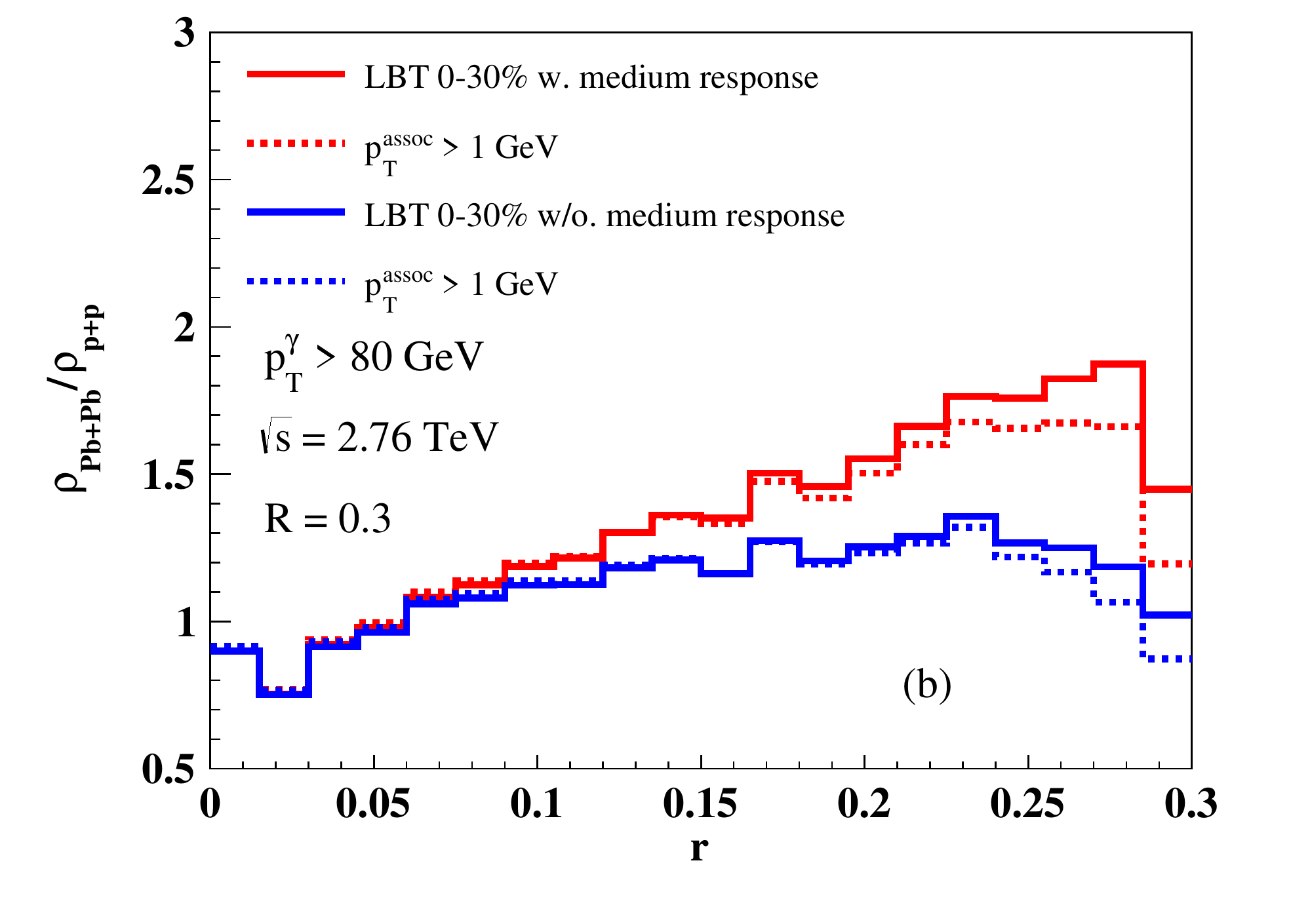}
\caption{(Color online) (a) The ``inclusive" transverse jet profile of $\gamma$-jets in central (0--30\%) Pb+Pb (red) and p+p collisions (blue) from LBT simulations and (b) their ratio with (red) and without (blue) contributions from jet-induced medium response at $\sqrt{s}=2.76$ TeV. The black lines in (a) are the contributions to the jet transverse profile from jet-induced medium response (recoil and ``negative" partons). Dashed lines are the jet profiles and their ratios with transverse momentum cut $p_T^{\rm asso}>1$ GeV for the associated partons.}
\label{profile2}
\end{figure}

In another definition which is used by the CMS experiment \cite{Chatrchyan:2013kwa}, all particles within a circular jet cone around the center of the jet are used to calculate the transverse energy profile which we refer to as ``inclusive''. These particles include not only those from the list of particles associated with the jet in FASTJET but also those that are not.  The jet profiles in p+p (red) and central 0-30\% Pb+Pb collisions (red), shown in Fig.~\ref{profile2}(a), are similar to the ``exclusive" jet profile in Fig.~\ref{profile1}(a) over most of the jet cone except the last couple of bins close to the edge of the cone ($r=R$) where the ``inclusive" jet profile is higher due to additional partons not in the list of particles associated with the jet in FASTJET.  Since the jet areas in FASTJET are irregular (not a perfect circle) some of the partons toward the edge of the jet area in the FASTJET fall outside of the circular cone used to calculate the jet profile in Eq.~(\ref{eq:profile}). In addition, the anti-$k_T$ algorithm in FASTJET may also prefers jets with sharp edges. This is why the jet profiles in both cases fall off more rapidly toward the edge of the jet cone. The effect of these additional partons (not in the list of particles associated with the jet in FASTJET) is more prominent in p+p collisions. The enhancement of the ``inclusive'' jet profile at large radius in central Pb+Pb collisions due to jet-medium interaction as shown in Fig.~\ref{profile2}(b) is slightly smaller than the ``exclusive" jet profile in Fig.~\ref{profile1}(b). In Fig.~\ref{profile2}, we also show jet profiles in p+p and Pb+Pb collisions and their ratios as dashed lines that exclude partons with transverse momentum $p_T^{\rm asso}<1$ GeV. This transverse momentum cut decreases the jet profiles and the enhancement in Pb+Pb collisions slightly toward the edge of the jet cone.

\begin{figure}[!htb]
\centering
\vspace{-0.6in}
\includegraphics[width=6.7cm,bb=15 15 500 500]{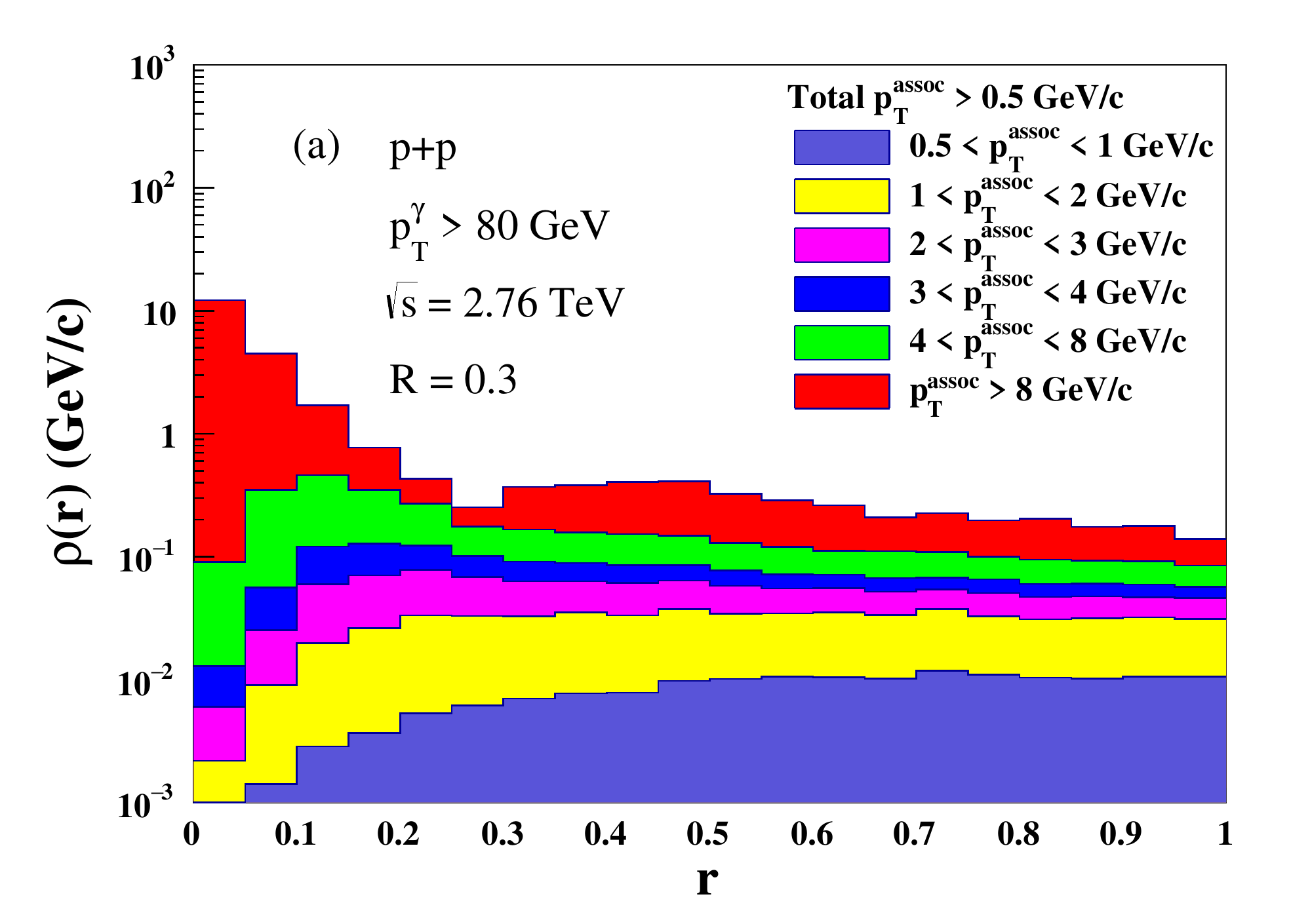} \\
\vspace{-0.63in}
\includegraphics[width=6.7cm,bb=15 15 500 500]{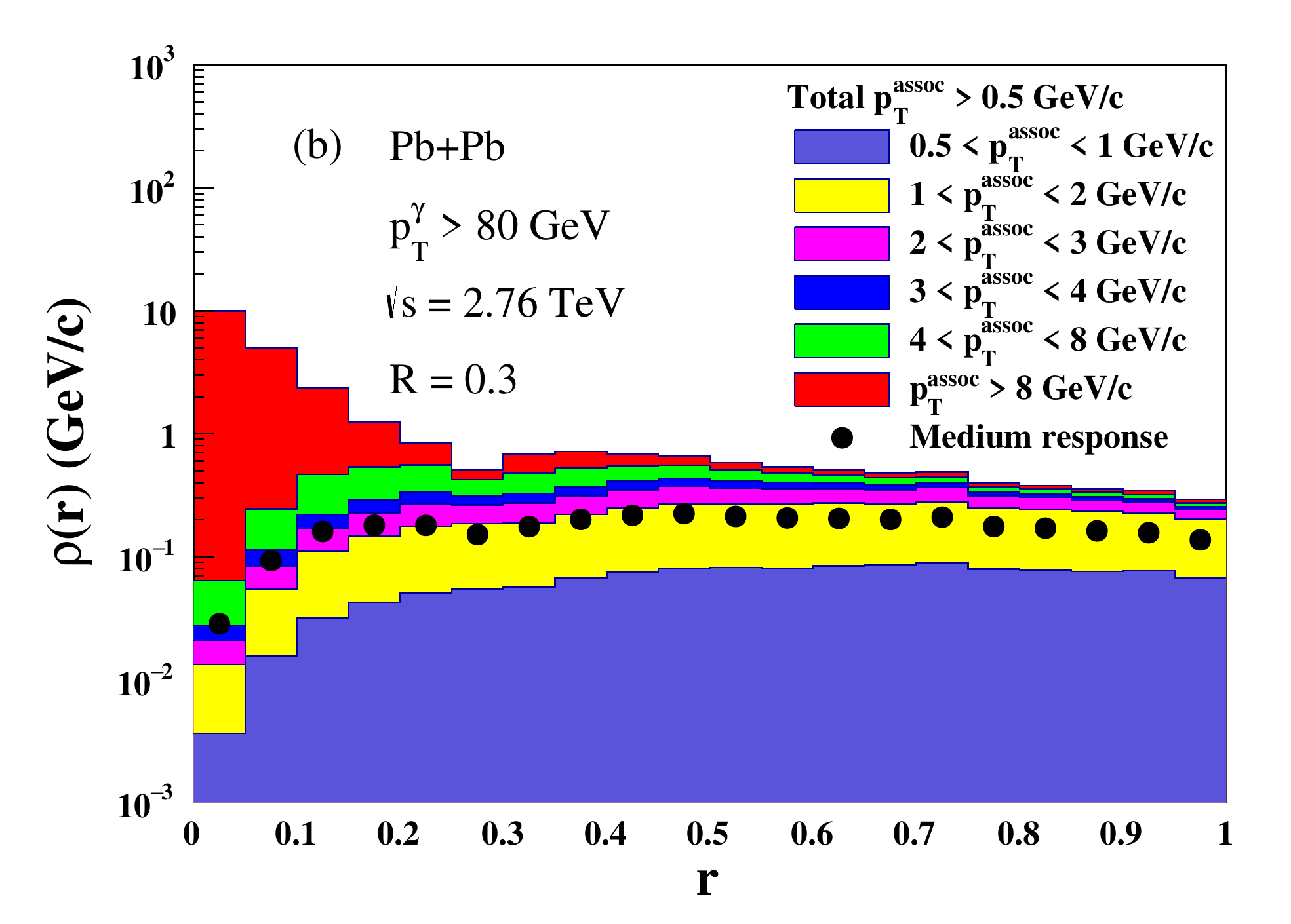}
\caption{(Color online) (a) Extended transverse jet profile of $\gamma$-jets in p+p and (b) central (0--30\%) Pb+Pb collisions at $\sqrt{s}=2.76$ TeV from LBT simulations. The solid circles in the lower panel show contributions from jet-induced medium response (recoil and ``negative" partons).}
\label{rho}
\end{figure}

\begin{figure}[!htb]
\centering
\vspace{-0.5in}
\includegraphics[width=6.7cm,bb=15 15 500 500]{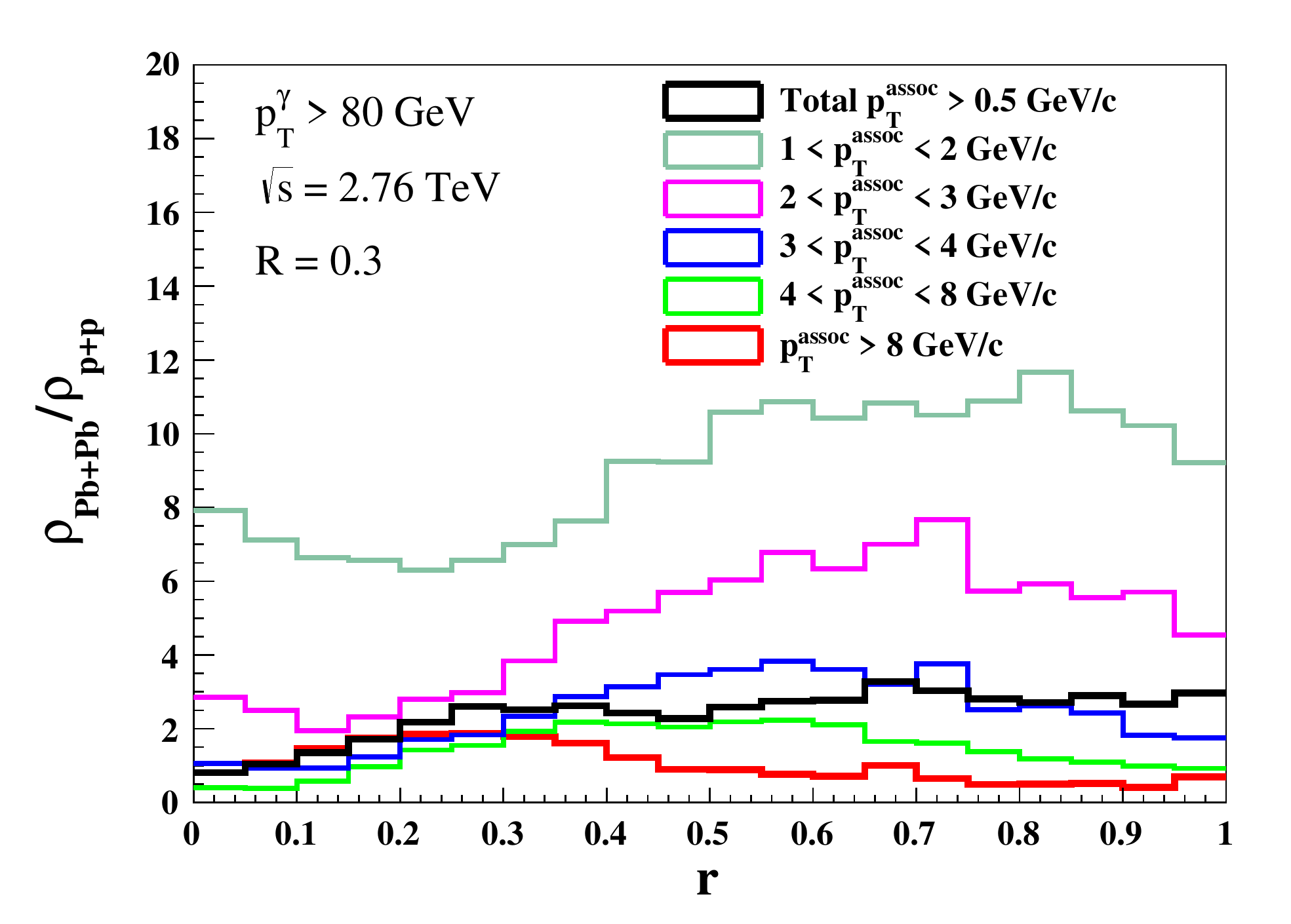}
\caption{(Color online) The ratio of transverse jet profile of $\gamma$-jets in central (0--30\%) Pb+Pb to that in p+p collisions at $\sqrt{s}=2.76$ TeV from LBT simulations.}
\label{rho-ratio}
\end{figure}

To investigate in more detail how the energy is transported inside and outside the jet-cone, we also calculate the extended transverse profile of the $\gamma$-jet  as shown in Fig.~\ref{rho} for $p_T^\gamma>80$ GeV/$c$ in p+p (upper panel) and 0-30\% central Pb+Pb (lower panel) collisions at $\sqrt{s}=2.76$ TeV. We use the similar method introduced by CMS for the studies of di-jets~\cite{Khachatryan:2016tfj}. With selected $\gamma$-jets using anti-$k_T$ algorithm with cone size $R=0.3$,  we extend the radius $r$ in the numerator of the ``exclusive" jet transverse profile in Eq.~(\ref{eq:profile})  to $ r=1$ and plot contributions from partons in different range of transverse momentum. We also plot the ratio of jet transverse profile in Pb+Pb over that in p+p collisions in Fig.~\ref{rho-ratio}. In the calculation of the extended jet profile, we have followed the scheme of underlying event background subtraction as used by CMS experiment \cite{Khachatryan:2016erx}. In this scheme, we take a side-band over the range $1.5<| \eta-\eta_{\rm jet}|<3.0$ in the event-averaged transverse energy distribution with respect to the center of the reconstructed jet as the background and subtract this $\phi$-dependent background from the jet profile for each of the $p_T$ range.

Within the LBT model, the suppression of the transverse energy within the very core $r<0.03$ inside the jet as seen in Figs.~\ref{profile1} and \ref{profile2} is caused by the suppression of energetic partons at $p_T>4$ GeV/$c$.  The enhancement of transverse energy toward the edge of the jet cone extends to the outside of the jet-cone at very large angle and are carried mostly by soft partons with $p_T<3$ GeV/$c$. In the lower panel of Fig.~\ref{rho} we also plot contributions from the recoil medium partons to the $\gamma$-jet profile in Pb+Pb collisions as solid circles. We can see that the enhanced transverse energy around the edge and outside the jet-cone is carried mostly by recoil medium partons from jet-medium interaction within the LBT model calculation.

\begin{figure}[!htb]
\centering
\vspace{-0.5in}
\includegraphics[width=6.7cm,bb=15 15 500 500]{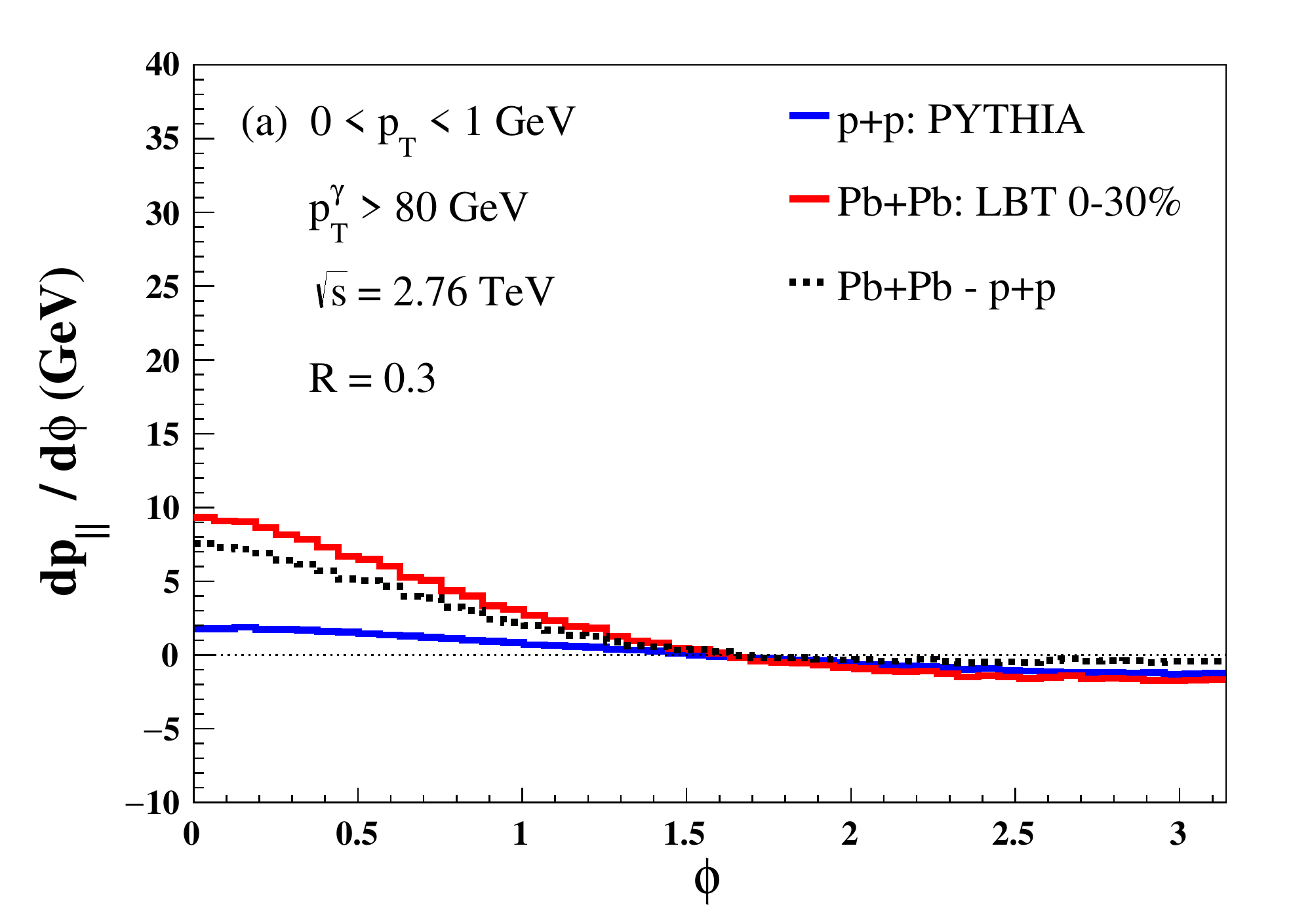} \\
\vspace{-0.73in}
\includegraphics[width=6.7cm,bb=15 15 500 500]{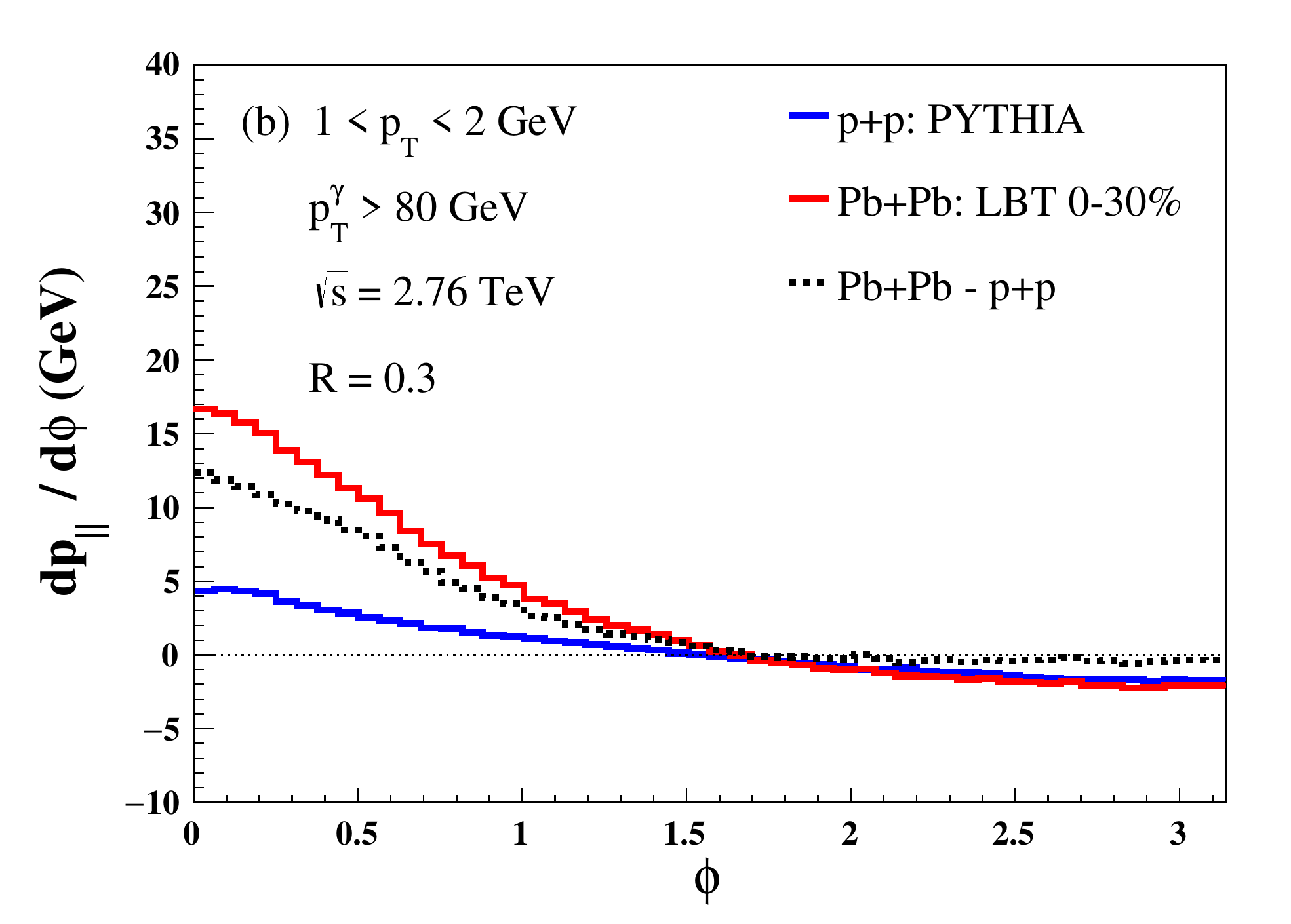} \\
\vspace{-0.73in}
\includegraphics[width=6.7cm,bb=15 15 500 500]{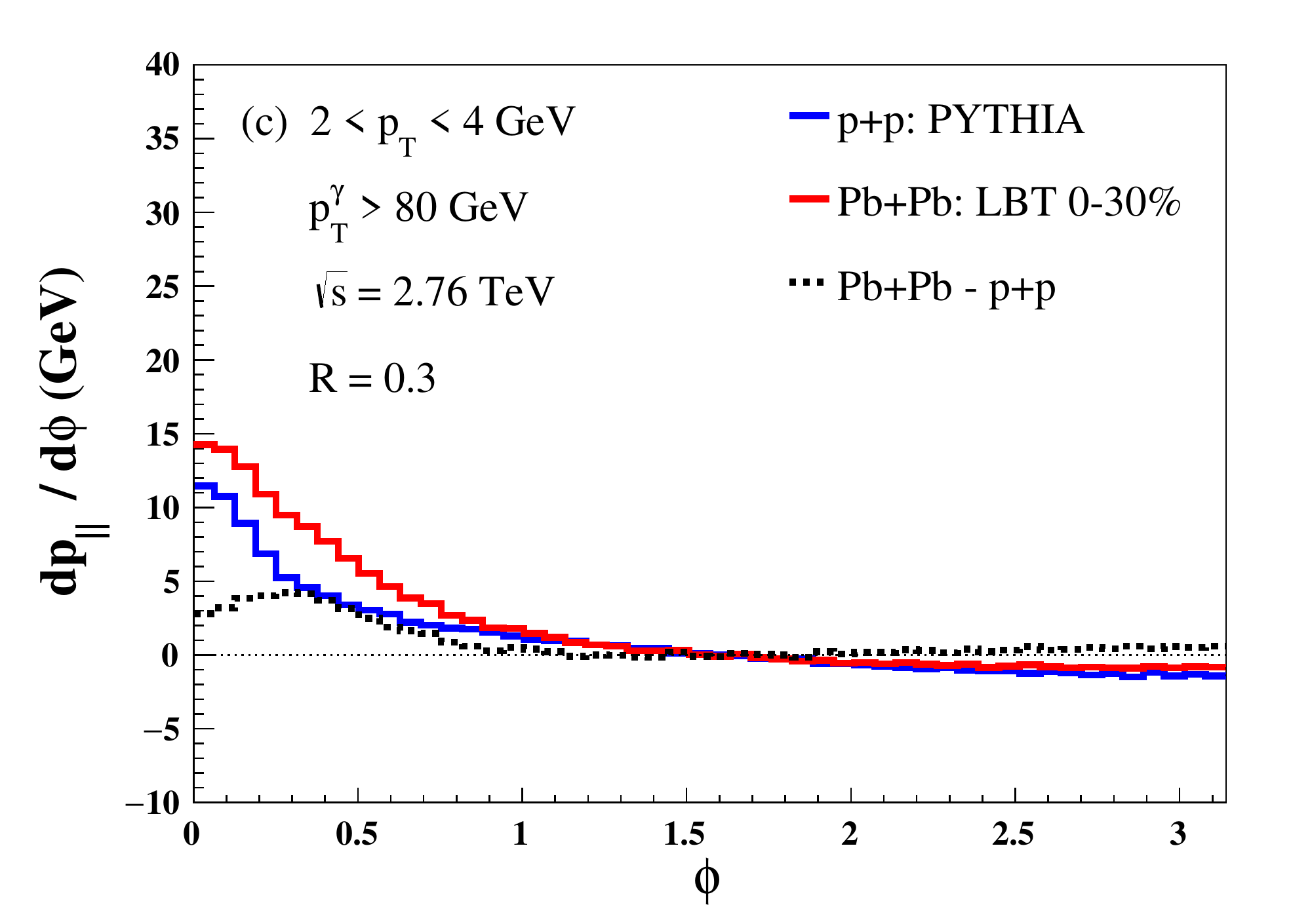} \\
\vspace{-0.73in}
\includegraphics[width=6.7cm,bb=15 15 500 500]{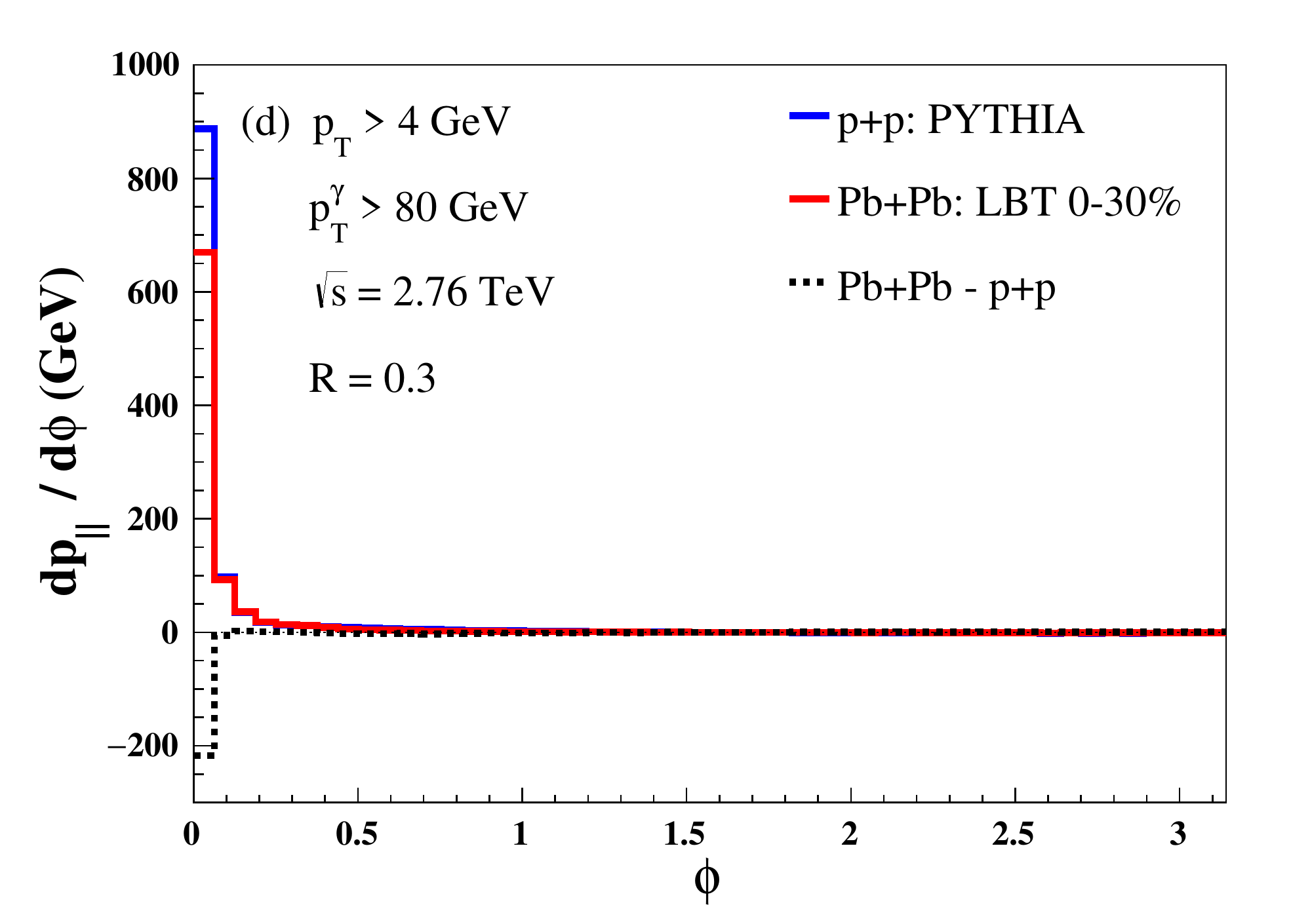}
\vspace{0.0 in}
\caption{(Color online) Azimuthal angle distributions of the parallel momentum flow along the jet direction from partons with different $p_T$ range in central (0--10\%) Pb+Pb (red solid), p+p collisions (blue solid) and their difference (black dashed) at $\sqrt{s}=2.76$ TeV from LBT simulations.}
\label{energyflow}
\end{figure}

As another illustration of the transport of transverse energy lost by leading jet shower partons in the medium, we plot in Fig.~\ref{energyflow} the energy flow,
\begin{equation}
p_\parallel=\sum_{\rm assoc} p_T^{\rm assoc}\cos(\phi-\phi_\gamma-\pi).
\end{equation}
which is defined as the sum of associated particles' momenta projected along the opposite direction of the direct photon for different range of the parton's
transverse momentum in p+p and 0-30\% central Pb+Pb collisions at $\sqrt{s}=2.76$ TeV. As we see from the LBT results, energetic partons at high $p_T$ has
a very narrow azimuthal angle distribution and are suppressed in Pb+Pb relative to p+p collisions. The energy lost by these leading jet shower partons is then carried by radiated and recoil soft partons. Recoil partons from jet-induced medium excitation contribute to most of the energy excess carried by soft partons in the jet direction that has a broad azimuthal angle distribution. 

\section{Summary and Discussions}
\label{summary}

Within the LBT Monte Carlo model for jet propagation in hot QGP medium, we have studied $\gamma$-jet production and modification in high-energy heavy-ion collisions at LHC. With a fixed effective value of the strong coupling $\alpha_{\rm s}$ within the model, we are able to describe the experimental data on $\gamma$-jet asymmetry, jet survival probability and azimuthal angle correlation well. We focused the study on the role of multiple jets in $\gamma$-jet correlations in high-energy heavy-ion collisions. Multiple jets associated with direct photon production through higher order QCD processes are found to have appreciable contributions to the inclusive $\gamma$-jet yield at low $p_T^{\rm jet}<p_T^\gamma$. They are the dominant contributions at large angle in the $\gamma$-jet azimuthal correlation. Jet-medium interaction suppresses not only the leading jet but also sub-leading jets in events with multiple jets in association with direct photon production. 

The suppression of sub-leading jets leads to the reduction of the inclusive $\gamma$-jet yields at lower $p_T^{\rm jet}<p_T^\gamma$ and also at the large azimuthal angle. This effectively narrows the $\gamma$-jet azimuthal angle correlation instead of broadening as one would have expected due to jet-medium interaction. The azimuthal angle correlation between the leading jet and the photon is almost the same in Pb+Pb and p+p collisions according to LBT results because of the dominance of the Sudakov form factor in $\gamma$-jet correlation from soft gluon radiation in large $p_T$ hard processes. This will pose a challenge for using $\gamma$-jet azimuthal correlation to study medium properties via large angle parton-medium interaction. 

The transverse profile of $\gamma$-jet with fixed $p_T^\gamma$ within the jet-cone is found to be broadened due to energy transport by the radiated gluons and recoil medium partons. These partons are relative soft compared to leading jet shower partons at the very core of the jet which is slightly suppressed due to parton energy loss. The energy carried by recoil medium partons is found to be transported to the outside of the jet-cone. They lead to the enhancement of the extended jet transverse profile both inside and outside the jet-cone. We also studied the energy flow in azimuthal angle relative to the direction of prompt photon. While the energy flow carried by large $p_T$ partons is suppressed within a narrow jet-cone, energy carried by soft partons is enhanced in the direction of the jet and the enhancement has a broad angle distribution due to jet-induced medium response. 

We have not included the hadronization of jet shower and recoil partons in this study within the LBT model. The hadronization will add an uncertainty to the jet energy scale in the order of 1 GeV.  It will also introduce some smearing in the jet transverse profile. The energy flow and jet profile analyses according to final particle's transverse momentum are only illustrative. One has to include the hadronization in order to provide a quantitative analysis according to final hadrons' transverse momenta.

\begin{acknowledgments}
We would to thank Wei Chen and Longgang Pang for providing the 3+1D hydro profiles used in this study. This work is supported in part by the National Science Foundation of China  under Grant No. 11221504, the Major State Basic Research Development Program in China (No. 2014CB845404), by the Director, Office of Energy Research, Office of High Energy and Nuclear Physics, Division of Nuclear Physics, of the U.S. Department of Energy under Contract Nos. DE-AC02-05CH11231 and DE-SC0013460, and by the US National Science Foundation within the framework of the JETSCAPE collaboration, under grant number ACI-1550228 and ACI-1550300.
\end{acknowledgments}

\end{document}